\documentclass[11pt, letterpaper]{article}

\usepackage{amsfonts} 
\usepackage{amsmath} 
\usepackage{mathrsfs}  %
\usepackage{amssymb}
\usepackage{empheq}
\usepackage{bm}
\usepackage{mathtools}
\usepackage{slashed}
\usepackage{relsize}
\usepackage{setspace}   
\usepackage{color}  
\usepackage[utf8,applemac]{inputenc} 
\usepackage{cite}
\usepackage{tikz}
\usetikzlibrary{calc, arrows.meta}
\usepackage{graphicx}
\usepackage{subfig}
\usepackage[normalem]{ulem}
\usepackage{hyperref}
\usepackage{comment}

\usepackage[toc,page]{appendix}

\usepackage[hang,flushmargin]{footmisc}

\usepackage[]{titlesec}
\titlespacing*{\section}{0pt}{0.5em}{-.5em}
\titlespacing*{\subsection}{0pt}{0.5em}{-1em}
\titlespacing*{\subsubsection}{0pt}{1em}{-1em}

\numberwithin{equation}{section}
\usepackage{dcolumn}
\usepackage{bm}
\usepackage[toc,page]{appendix}

\usepackage{vmargin}
\setmargrb{2cm}{1cm}{2cm}{2cm}

\usepackage{multirow}

\usepackage{enumitem}

\setlist{
  topsep = 0em,
  }

\newcommand{\dd}{\text{d}}

\newcommand\nts{\negthickspace}

\newcommand\defeq{\mathrel{\mathop:}=}

\newcommand\LL{\mathcal{L}}
\newcommand\VV{\mathcal{V}}
\newcommand\UU{\mathcal{U}}
\newcommand\RR{\mathcal{R}}

\newcommand\EE{\mathcal{E}}

\newcommand\DD{\mathcal{D}}
\newcommand\OO{\mathcal{O}}
\newcommand\CC{\mathcal{C}}
\newcommand\NN{\mathcal{N}}
\newcommand\MM{\mathcal{M}}
\newcommand\BB{\mathcal{B}}
\renewcommand\AA{\mathcal{A}}

\newcommand\PP{\mathcal{P}}

\definecolor{bluegreen}{RGB}{0,102,102}

\def\oeq{\stackrel{\circ}{=}}

\usepackage{pict2e}
\makeatletter
\newcommand{\loplus}{\mathbin{\mathpalette\dog@lsemi{+}}}
\newcommand{\dog@lsemi}[2]{\dog@semi{#1}{#2}{270,90}}
\newcommand{\dog@semi}[3]{%
  \begingroup
  \sbox\z@{$\m@th#1#2$}%
  \setlength{\unitlength}{\dimexpr\ht\z@+\dp\z@\relax}%
  \makebox[\wd\z@]{\raisebox{-\dp\z@}{%
    \begin{picture}(1,1)
    \linethickness{\variable@rule{#1}}
    \roundcap
    \put(0.5,0.5){\makebox(0,0){\raisebox{\dp\z@}{$\m@th#1#2$}}}
    \put(0.5,0.5){\arc[#3]{0.5}}
    \end{picture}%
  }}%
  \endgroup
}
\newcommand{\variable@rule}[1]{%
  \fontdimen8  
  \ifx#1\displaystyle\textfont3\else
    \ifx#1\textstyle\textfont3\else
      \ifx#1\scriptstyle\scriptfont3\else
        \scriptscriptfont3\relax
  \fi\fi\fi
}
\makeatother

\hypersetup{
    colorlinks,%
    citecolor=blue,%
    filecolor=blue,%
    linkcolor=blue,%
   urlcolor=cornflower,
   linktoc=page
}

\definecolor{scarlet}{rgb}{0.8, 0, 0}
\definecolor{brick}{rgb}{0.64314, 0, 0}
\definecolor{cornflower}{rgb}{0.12549, 0.29020, 0.52941}

\def\mathcolor#1#{\@mathcolor{#1}}
\def\@mathcolor#1#2#3{%
  \protect\leavevmode
  \begingroup
    \color#1{#2}#3%
  \endgroup
}

\begin{document}

\title{\Large{\textbf{\sffamily The symplectic potential for leaky boundaries}}}
\author{\sffamily Robert McNees${}^1$ and C\'eline Zwikel${}^2$
\date{\small{\textit{
$^1$Department of Physics, Loyola University Chicago,
Chicago, IL, USA\\
$^2$Perimeter Institute for Theoretical Physics,\\ 31 Caroline Street North, Waterloo, Ontario, Canada N2L 2Y5\\}}}}

\maketitle

\begin{abstract}
Charges associated with gauge symmetries are defined on boundaries of spacetimes. But these constructions typically involve divergent quantities when considering asymptotic boundaries. Different prescriptions exist to address this problem, based on ambiguities in the definition of the symplectic potential. We propose a method well suited to leaky boundaries, which describe spacetimes than can exchange matter or radiation with their environment. The main advantage of this approach is that it relies only on the bulk Lagrangian and it is not tied to a specific choice of boundary conditions. The prescription is applied to four dimensional Einstein-Hilbert gravity in the partial Bondi gauge. This leads to a finite symplectic potential for unconstrained boundary data and reveals two new corner symplectic pairs associated with the relaxation of the gauge. 
\end{abstract}

\tableofcontents

\section{Introduction}
In recent years, there has been growing interest in open systems in gravity, which are spacetimes or regions of spacetimes with leaky boundaries. Leakage in these systems can occur due to some flux passing through the boundary or simply because the system is coupled in an unspecified way to an external environment. 
 
A familiar description of an open system in gravity is the Bondi mass loss formula for asymptotically flat spacetimes \cite{Trautman:1958zdi, Bondi:1962px}. For example, gravitational waves produced through black hole mergers or some other process carry energy to the null asymptotic boundary of spacetime. Likewise, one can define non-reflective boundary conditions in asymptotically anti-de Sitter spacetimes that model the exchange of energy with an external system, leading to a similar result for large black holes which might otherwise be in equilibrium \cite{Ashtekar:2015lla, Compere:2019bua, Compere:2020lrt}. In both cases, conservation of a charge (the mass) is replaced by a flux-balance law that describes the leakage or exchange. Open systems have also been considered in lower dimensions. A simple toy model consisting of Jackiw-Teitelboim gravity (a dilaton gravity theory in two dimensions) coupled to a massless scalar field was presented in \cite{Grumiller:2023ahv}. This additional scalar field models the radiation emitted by a black hole. In this setup, where there is no coupling to an external thermal bath, the system exhibits leakage. Another example of a leaky system is the Einstein-Maxwell theory in three dimensions, which can accommodate electromagnetic radiation \cite{Bosma:2023sxn}. Additional details and examples may be found in the review \cite{Fiorucci:2021pha}.

The boundary of a finite region of spacetime typically exhibits leaky characteristics rather than preserving a closed system. A typical example is a black hole horizon, which can both evaporate or absorb matter and radiation \cite{Donnay:2016ejv, Chandrasekaran:2018aop, Adami:2020amw, Adami:2021sko, Adami:2021nnf, Freidel:2022vjq, Adami:2024gdx, Odak:2023pga, Ciambelli:2023mir}. In this paper the focus is on leaky boundaries ``at infinity.'' These asymptotic boundaries are represented by a component of the boundary of a regulated (finite) region of spacetime. Removing the regulator involves an infinite limit in which quantities like the symplectic potential, or charges constructed from the symplectic current, may exhibit divergences. In that case, one needs a procedure for addressing these divergences and obtaining finite quantities.

A prescription for obtaining a finite symplectic potential, current, and charges at infinity was introduced in \cite{McNees:2023tus}. An important feature of this approach, which is based on an inherent ambiguity in the definition of the symplectic potential, is that it relies only on information in the bulk Lagrangian and is not tied to a specific choice of boundary conditions. This makes it particularly well suited for applications to open systems and leaky boundaries, and a useful alternative to prescriptions that incorporate a specific set of boundary and corner terms in the action  \cite{Compere:2008us, Detournay:2014fva, Compere:2020lrt, Freidel:2020xyx, Fiorucci:2020xto, deHaro:2000xn, Chandrasekaran:2021vyu, Bianchi:2001kw, Freidel:2021cjp, Freidel:2021fxf, Margalef-Bentabol:2020teu, G:2021xvv, Margalef-Bentabol:2022zso,Capone:2023roc}.\footnote{An alternative way of dealing with leaky boundaries is to add external degrees of freedom known as ``edge modes,'' that effectively restore gauge invariance and close the system \cite{Donnelly:2016auv,Speranza:2017gxd,Geiller:2019bti,Ciambelli:2021nmv,Freidel:2021dxw,Adami:2024gdx}.} The procedure was given in general terms in \cite{McNees:2023tus}, then applied to gravitational theories in two and three dimensions. Here, we apply it to gravity in four dimensions and fully account for the presence of radiating degrees of freedom. The extension to dimension $d+1 > 4$ and for adding matter is straightforward.\,\footnote{A similar construction for spin-one fields is studied in \cite{Freidel:2019ohg,Campoleoni:2023eqp}.}

In this paper we apply this prescription in the partial Bondi gauge (PBG), which is a useful framework for describing asymptotic boundaries (whether null, timelike, or spacelike) of four-dimensional solutions of Einstein gravity with or without a cosmological constant. The PBG geneneralizes the familiar Bondi-Sachs \cite{Bondi:1962px,Sachs:1962zza} and Newman-Unti \cite{Newman:1962cia} gauges in two ways. First, it is a partial gauge fixing that does not impose a condition on the radial coordinate. This allows one to perform calculations in PBG and then specialize to Bondi-Sachs or Newman-Unti gauge afterwards, or easily translate results between these two gauges. It has been shown that the relaxed condition on the radial coordinate leads to additional symmetries \cite{Geiller:2024amx} which are not yet fully understood. Second, the PBG allows for the most general boundary data compatible with the equations of motion and conformal compactification, without any \emph{ad hoc} fixing. In particular, we allow for arbitrary time dependence and unconstrained variations of the boundary data. This encompasses a number of solutions of Einstein's equations, such as gravitational shock waves \cite{He:2023qha,He:2024vlp} and Robinson-Trautman spacetimes \cite{Robinson:1962zz}, and accommodates proposals for radiation in anti-de Sitter or de Sitter spacetimes \cite{Compere:2019bua, Compere:2020lrt, Bonga:2023eml, Ciambelli:2024kre, Fernandez-Alvarez:2021yog}. Techniques similar to the one presented here have been used to address divergent quantities at infinity for PBG with some additional restrictions \cite{Compere:2018ylh, Geiller:2024amx, Geiller:2024ryw}. See also the recent work in \cite{Riello:2024uvs}.

Our main results are the application of the method described in \cite{McNees:2023tus} to Einstein-Hilbert gravity in four dimensions (section \ref{sec:PrescriptionForEH}) and a detailed expression for the resulting symplectic potential as a function of boundary data and dynamical quantities in PBG (section \ref{sec:finitesympl}). This latter result is given for spacetimes with a component $\BB$ of the asymptotic boundary obtained as the $r \to \infty$ limit of a timelike surface. This surface may have corners $\partial \BB$ where it intersects other components of the boundary. The $r\to\infty$ divergences in this potential are either $\delta$-exact ($\Lambda \neq 0$) or vanish entirely ($\Lambda = 0$), so that the symplectic current $\omega$ and associated codimension-2 form $k$ are manifestly finite. We identify the symplectic pairs appearing in the potential, including new pairs with support on corners, and demonstrate a discontinuity in the flat limit $\Lambda \to 0$.

The outline of the paper is as follows. In section \ref{sec:Prescription} we review the prescription for the symplectic potential given in \cite{McNees:2023tus}. This includes a choice of a residual ambiguity that corresponds to a finite corner contribution at infinity. Section \ref{sec:PartialBondiGauge} gives a brief review of the partial Bondi Gauge, then describes ``partially on-shell'' conditions for the fields which enforce a subset of the equations of motion. In section \ref{sec:SymplecticPotentialInBondiGauge} we apply the prescription to the Einstein-Hilbert Lagrangian and obtain a new symplectic potential $\widetilde{\Theta}$ that resolves an obstruction to defining charges. The finite part of $\widetilde{\Theta}$ is presented in a compact form which identifies the symplectic pairs. We conclude with a brief discussion summarizing our results, commenting on open questions, and highlighting some work that will appear in an upcoming paper. More details on the definition of charges for partially on-shell fields can be found in appendix \ref{app:DiffeoCharges}, complete results for the solution space in PBG are given in appendix \ref{app:EOMPBG}, and useful identities are compiled in appendix \ref{app:Identities}.

Many calculations in this paper were carried out using the xAct suite of packages for the Wolfram Language \cite{Martin-Garcia-xAct, Martin-Garcia:2008ysv, Brizuela:2008ra, Nutma:2013zea}. A \textsc{Mathematica} notebook with the full expression for $\widetilde{\Theta}_{0}^{r}$ is available as an ancillary file alongside the ar$\chi$iv upload. Conventions not explicitly given in the text, along with a number of useful results, may be found at \cite{McNees-Useful} and in appendix \ref{app:Identities}.

\section{Prescription for the presymplectic potential}
\label{sec:Prescription}

In this section we review a method for obtaining a finite symplectic potential (up to $\delta$-exact terms), current, and associated codimension-2 form ``at infinity.'' The construction is first described in general terms and then applied to the Einstein-Hilbert Lagrangian in four dimensions. This approach was presented in \cite{McNees:2023tus}, where it was applied to models in two and three dimensions. An important difference between those examples and the application considered here is the presence of propagating degrees of freedom. This was briefly discussed in \cite{McNees:2023tus} but is carefully accounted for here.

Before discussing our prescription in detail, it is useful to review a few facts about the symplectic potential. First, since the potential appears as a total derivative in the variation of a bulk Lagrangian $L_{M}$, it is only defined up to the equivalence
\begin{gather}\label{eq:BasicEquivalence}
    \Theta \sim \Theta + \dd \vartheta ~.
\end{gather}
Shifting the potential by the closed form $\dd\vartheta$ has no impact on the equations of motion, but it may affect the charges obtained with the associated symplectic current, as well as the variational formulation of the theory. One can also add a surface term $\dd L_{\partial M}$ to the bulk Lagrangian without affecting the equations of motion, which will change the potential by a total variation $\delta L_{\partial M}$. Surface terms are usually added to ensure a well-defined variational formulation of the theory for a specific set of boundary conditions, and may be necessary for constructions involving a boundary stress tensor. Depending on the conditions satisfied by the fields on $\partial M$, they might also lead to a term like \eqref{eq:BasicEquivalence}, but here we make no such assumptions and treat $\delta L_{\partial M}$ and $\vartheta$ as distinct. In that case this sort of modification should not affect the definition of charges via the symplectic current $\omega$, which is insensitive to $\delta$-exact terms in $\Theta$.

A number of prescriptions for working with symplectic potentials have been introduced for different physical systems. Typically, one begins with $\Theta$ obtained from straightforward variation of a bulk Lagrangian, then identifies a $\Theta + \dd\vartheta + \delta L_{\partial M}$ that satisfies requirements like finiteness in a certain limit, and a particular functional dependence on the fields on $\partial M$. The first condition is important for charges defined at infinity, since the relevant quantities might not exist in the implied limit \cite{Regge:1974zd, Balasubramanian:1999re, Emparan:1999pm, Hollands:2005wt, Hollands:2005ya, Mann:2005yr, Mann:2006bd, Mann:2008ay}. In the case of a closed system, the second requirement is usually because one wishes to impose a specific set of boundary conditions. Several prescriptions relying on boundary Lagrangians have been given for this purpose \cite{Compere:2008us, Detournay:2014fva, Grumiller:2019fmp, Compere:2020lrt, Freidel:2020xyx, Fiorucci:2020xto, deHaro:2000xn, Chandrasekaran:2021vyu, Bianchi:2001kw, Freidel:2021cjp, Freidel:2021fxf, Margalef-Bentabol:2020teu, G:2021xvv, Margalef-Bentabol:2022zso,Capone:2023roc}. However, open (leaky) systems may require a more flexible approach that can accommodate coupling to as-yet-unspecified external systems. And, as mentioned above, the definition of charges via a symplectic current $\omega$ should be insensitive to contributions from a boundary Lagrangian. It would therefore be useful to have a procedure for defining charges which is not tied to a specific choice of boundary conditions.

The approach presented here has the benefit of relying only on the equivalence \eqref{eq:BasicEquivalence}. It eliminates an obstruction to defining charges at infinity, leads to a finite symplectic potential and current, and does not depend on a particular set of surface or corner terms in the action. Furthermore, the part of $\vartheta$ responsible for these features is determined entirely from information present in the bulk Lagrangian.\footnote{A residual ambiguity in the definition of $\vartheta$ is addressed in more detail in the next subsection, and in section \ref{sec:Discussion}.} This disentangles the construction of charges from the process of specifying boundary conditions, and is applicable to leaky systems.

\subsection{General Construction}\label{sec:GeneralConstruction}
Consider a theory with fields $\Psi_{i}$ on a four-dimensional spacetime $M$, described by a Lagrangian $L_{M}(\Psi)$. Here and elsewhere, dependence on $\Psi$ typically means a function of both the fields and their derivatives. Under a variation of the fields, the change in the Lagrangian is
\begin{gather}
    \delta L_{M}(\Psi) = E^{i}(\Psi)\,\delta\Psi_{i} + \dd \Theta(\Psi;\delta \Psi) ~.
\end{gather}
The $E^{i}(\Psi)$ are the equations of motion, and $\Theta$ is a (pre)symplectic potential that depends on the fields and is linear in the field variations. Given a potential $\Theta$, the associated current $\omega$ is obtained via the antisymmetrized second variation
\begin{gather}
    \omega(\Psi; \delta_1 \Psi, \delta_2 \Psi) = \delta_{2} \Theta(\Psi;\delta_1 \Psi) - \delta_{1} \Theta(\Psi;\delta_2 \Psi) ~.
\end{gather}
When the field variations satisfy the linearized equations of motion, $\delta (E^{i}) =0$, we have $\dd\omega = 0$ and there exist codimension-2 forms $k$ such that $\omega = \dd k$. For a gauge symmetry $\delta_{\xi} \Psi$ of the Lagrangian, the contraction of $k$ with $\xi$ is defined as
\begin{gather}
    \dd k_{\xi} = \omega_{\xi} = \omega(\Psi; \delta\Psi, \delta_{\xi}\Psi ) ~.
\end{gather}
This can be integrated over a codimension-2 surface $\CC$ to obtain charges 
\begin{gather}\label{eq:PreCharge}
    \delta\!\!\!/ Q_{\xi} = \int_{\CC} k_{\xi} ~.
\end{gather}
In general, charges defined in this manner may not be integrable, as indicated by the $\delta\!\!\!/$ notation. But before the question of integrability can be addressed, one must establish whether or not the quantities described above actually \textit{exist}. This is an issue when $\CC$ is a cut of infinity, since the quantities leading up to \eqref{eq:PreCharge} may diverge in that case.

For a spacetime $M$ that includes an asymptotic region ``at infinity,'' we represent this region as the $r \to \infty$ limit of an isosurface $\BB$ of some scalar field $r$. Generally speaking, quantities defined at infinity must be understood as the $r \to \infty$ limit of quantities evaluated at finite $r$. At large $r$, let $\CC$ be a closed codimension-2 surface given by the intersection of $\BB$ with another surface $\NN$, with $\NN$ described on a neighborhood of $\BB$ as an isosurface of another scalar field $u$. Then the charges described above should be obtained from
\begin{gather}\label{eq:ChargesWithLimit}
    \delta\!\!\!/ Q_{\xi} = \lim_{r \to \infty} \int_{\CC} k_{\xi}^{ur} ~.
\end{gather}
It is easy to see why this limit may not exist. From $\dd k_{\xi} = \omega_{\xi}$, we have
\begin{gather}
    \partial_{r} k_{\xi}^{ur} + \partial_{A} k_{\xi}^{uA} = \omega_{\xi}^{u} ~,
\end{gather}
where the index $A$ denotes directions along $\CC$. Since $\CC$ is closed we can discard total derivatives $\partial_{A}$ to find
\begin{gather}\label{eq:Obstruction}
    \partial_{r} \int_{\CC} k_{\xi}^{ur} = \int_{\CC} \omega_{\xi}^{u} ~.
\end{gather}
It is clear that the limit in \eqref{eq:ChargesWithLimit} exists only if (the integral over $\CC$ of) $\omega_{\xi}^{u}$ falls off faster than $r^{-1}$. This may not be the case for a current obtained from an arbitrary potential $\Theta$. Indeed, it is often \emph{not} the case for $\Theta$ obtained directly from the variation of $L_{M}$.

One way to address this problem is to tighten the fall-off conditions on the fields and their variations so that $\omega_{\xi}^{u}$ has the desired fall-off. But this approach may exclude field configurations of interest. For example, in asymptotically flat gravity one would have to consider field variations that fall off more rapidly than the $1/r$ asymptotics of the Schwarzschild solution \cite{Regge:1974zd}. Another approach is to exploit the ambiguity \eqref{eq:BasicEquivalence} in $\Theta$, and work with a different potential $\widetilde{\Theta} = \Theta + \dd \vartheta$ such that the relevant component of $\tilde{\omega}$ has the desired behavior. Under the shift
\begin{gather}
    \Theta^{\mu} \to \widetilde{\Theta}^{\mu} = \Theta^{\mu} + \partial_{\nu}\vartheta^{\mu\nu} ~,
\end{gather}
the current changes as
\begin{gather}
    \omega^{\mu} \to \tilde{\omega}^{\mu} = \omega^{\mu} + \partial_{\nu}\Big(\delta_{2}\,\vartheta^{\mu\nu}(\Psi;\delta_1\Psi) - \delta_{1}\,\vartheta^{\mu\nu}(\Psi;\delta_2\Psi)\Big) ~.
\end{gather}
Our prescription is to take $\vartheta^{ur}$ such that
\begin{gather}
    \widetilde{\Theta}^{u} = \Theta^{u} + \partial_{r} \vartheta^{ur}  = 0 + o(r^{-1}) ~,
\end{gather}
with $o(r^{-1})$ indicating terms that fall off faster than $r^{-1}$. This choice is always possible as long as the fields are sufficiently smooth functions of $r$. Then $\tilde{\omega}^{u} = o(r^{-1})$, which ensures that the limit in \eqref{eq:ChargesWithLimit} exists and $\tilde{k}^{ur}$ is finite as $r \to \infty$.\,\footnote{Weaker conditions are possible, but will not be considered here. For example, it is only the integral of $\omega_{\xi}$ over $\CC$ that must fall off like $r^{-1}$. So terms in $\omega_{\xi}$ which vanish when integrated over $\CC$ could have any behavior at large $r$. Or, the full $\omega^{u}$ might include terms with some other behavior at large $r$, which vanish when one of the field variations is contracted with an asymptotic symmetry.} Upon making the shift described above, the $r$ component of $\widetilde{\Theta}$ is
\begin{gather}
    \widetilde{\Theta}^{r} = \Theta^{r} + \partial_{u}\,\vartheta^{ru} ~.
\end{gather}
Then $\partial_{\nu} \tilde{k}^{r\nu} = \tilde{\omega}^{r}$ implies that $\tilde{\omega}^{r}$ is finite as well, so that any $r\to \infty$ divergences in $\widetilde{\Theta}^{r}$ must be $\delta$-exact. Applying this construction to gravity in four dimensions, we find that $\widetilde{\Theta}^{r}$ is either finite ($\Lambda = 0$) or else finite up to $\delta$-exact terms ($\Lambda \neq 0$).

An essential point is that this procedure does not require the addition of a particular set of surface or corner terms to the bulk action. This is not an example of ``holographic renormalization'' via a boundary Lagrangian. We have simply rearranged the content of $\Theta^{\mu}$ in a manner that preserves the quantity $\partial_{\mu}\Theta^{\mu}$ appearing in $\delta L_{M}$. The result is a finite symplectic current, codimension-2 form, and charges using only information contained in the first variation of the bulk Lagrangian. This construction can be applied to a specific variational problem by supplementing the bulk action with whatever surface and corner terms are needed to implement a particular set of boundary conditions. Extracting such terms from $\widetilde{\Theta}$ will be explored in detail in an upcoming publication \cite{McNees:2025acf}.

Notice that this procedure does not completely fix the ambiguity in the potential: the quantity $\vartheta^{ur}$, and hence $\widetilde{\Theta}^{r}$, is only defined up to an $r$-independent piece. This residual ambiguity has two immediate consequences. First, it may affect the integrability of $\delta\!\!\!/Q_{\xi}$. In that case, one might choose the finite part of $\vartheta$ by demanding the existence of integrable charges for particular asymptotic symmetries, or to realize as many asymptotic symmetries as possible \cite{Troessaert:2013fma,Grumiller:2016pqb,Perez:2016vqo,Grumiller:2017sjh,Geiller:2021vpg,Ciambelli:2023ott}, or requiring covariance and stationarity \cite{Wald:1999wa,Chandrasekaran:2018aop,Odak:2022ndm,Rignon-Bret:2024gcx}. Second, a finite shift in $\widetilde{\Theta}^{r}$ may affect the surface and corner terms which must be added to the action to describe gravity with a specific set of boundary conditions. In the next section we apply our procedure to Einstein-Hilbert gravity in four dimensions, and make an explicit choice for the finite part of $\vartheta^{ur}$ motivated by the structure of $\delta$-exact terms in $\widetilde{\Theta}^{r}$ and how they impact the variational problem (an analysis we will return to in \cite{McNees:2025acf}). But it is possible that the two motivations outlined above require different choices for the finite part of $\vartheta$ .

As a last comment, we note that covariant phase space methods assume the fields are on-shell when evaluating the charges: $E^{i} = 0$. However, beginning in section \ref{sec:PartialBondiGauge} we only require the weaker condition $E^{i}\,\delta\Psi_i = 0$. Some field variations are assumed to vanish to preserve a (partial) choice of gauge in which some of the $\Psi_i$ are set to zero. The equations of motion conjugate to those variables, normally enforced as constraints, are not imposed. Appendix \ref{app:DiffeoCharges} discusses this in more detail and explains how the usual construction of charges is modified. Although we consider field configurations which are not fully on-shell or completely gauge-fixed, we still use terms like ``symplectic'' to describe the potential and associated current, and ``covariant phase space'' as a general description of the method used to obtain charges and related quantities.

\subsection{Prescription for Einstein-Hilbert gravity}
\label{sec:PrescriptionForEH}
In this section we apply the prescription outlined above to the Einstein-Hilbert Lagrangian in four dimensions, with or without a cosmological constant. We also make a specific choice for a finite corner contribution to the symplectic potential. 

The Einstein-Hilbert action, including a cosmological constant, includes the bulk integral
\begin{equation}\label{eq:EHBulk}
   \frac{1}{2\kappa^2}\int_{M} \nts d^4x \sqrt{-g}\left( R-2\Lambda\right) ~,
\end{equation}
with $\kappa^{2} = 8\pi G$. The first variation of the Lagrangian density is
\begin{equation}
    \delta L_{M} = E^{\mu\nu} \, \delta g_{\mu\nu}+ \partial_\mu\Theta^\mu ~,
\end{equation}
where $2\kappa^{2} E^{\mu\nu}$ is (minus) the Einstein tensor with cosmological constant, and $\Theta^{\mu}$ is a symplectic potential with components
\begin{gather}\label{def:symplpot}
    \Theta^{\mu} = \frac{1}{2\,\kappa^{2} }\,\left(\partial_{\nu} H^{\mu\nu} + \Gamma^{\mu}_{\nu\lambda} H^{\nu\lambda}\right) \,\,,\qquad 
    H^{\mu\nu} = \sqrt{-g}\,\left(\big(\delta g\big)^{\mu\nu} - g^{\mu\nu} \big(\delta g\big)^{\lambda}{}_{\lambda}\right) ~.
\end{gather}
The notation $(\delta g)^{\mu\nu}$ refers to the variation of the covariant components of the metric with indices raised by the inverse metric; this differs from the variation of the inverse metric by a minus sign: $\delta(g^{\mu\nu}) = - (\delta g)^{\mu\nu}$. We will sometimes refer to \eqref{def:symplpot}, obtained directly from the variation of $L_{M}$, as the bare symplectic potential.

We are interested in spacetimes with an asymptotic region ``at infinity,'' with topology $\mathbb R \times S^{2}$.\,\footnote{The generalization to $d+1$ dimensions, replacing $S^{2}$ with $S^{d-1}$ or some other closed $\CC$, is straightforward.} In practice we work on a regulated spacetime, also referred to as $M$, that is a region of the full spacetime. This regulated spacetime has boundary $\partial M$ which may include multiple components. One component of $\partial M$ is taken to be a constant $r$ surface $\BB$ such that the asymptotic region of the full spacetime is recovered in the limit $r \to \infty$. We assume that $\partial M$ includes corners, which we denote $\partial \BB$, where $\BB$ intersects another component of $\partial M$. The surface $\BB$ has coordinates $u$ generating $\mathbb R$ and $x^A$ spanning the $S^{2}$, which are collectively denoted $x^{i}$.

The $u$-component of the symplectic potential above is given by
\begin{gather}\label{eq:GeneralThetau}
    \Theta^{u} = \frac{1}{2 \kappa^{2}}\,\partial_{r}H^{ur} + \frac{1}{2\,\kappa^{2}} \left(\partial_{i} H^{ui} + \Gamma^{u}_{\nu\lambda}\,H^{\nu\lambda} \right) ~.
\end{gather}
This expression will simplify when we specialize to partial Bondi gauge, but for now we give general expressions which could also be used in other gauges (for instance, Fefferman-Graham gauge when $\Lambda \neq 0$, see appendix D of \cite{McNees:2023tus} or \cite{Ciambelli:2023ott} for another example in three dimensions). The procedure outlined in the previous section instructs us to shift to a symplectic potential with $\widetilde{\Theta}^{u} = \Theta^{u} + \partial_{r}\vartheta^{ur} = o(r^{-1})$, so we write \eqref{eq:GeneralThetau} as
\begin{gather}\label{eq:GeneralThetauShift}
    \Theta^{u} = \frac{1}{2 \kappa^{2}}\,\partial_{r} H^{ur} + \partial_{r} \left[ \, \frac{1}{2 \kappa^{2}} \int^{r} \nts d r'\,\left( \partial_{i} H^{ui} + \Gamma^{u}_{\nu\lambda}\,H^{\nu\lambda} \right)\right] + o(r^{-1}) ~.
\end{gather}
The $H^{ur}$ term already appears in $\Theta^{u}$ as a total $r$ derivative, and contributions from the other terms in \eqref{eq:GeneralThetau} are collected in the integral. Only the part of this integral which diverges in the $r \to \infty$ limit is relevant, since $o(r^{-1})$ terms in $\Theta^{u}$ may be neglected. Thus, we can shift from $\Theta$ to an appropriate $\widetilde{\Theta}$ by taking
\begin{gather}\label{eq:varthetapresc}
    \vartheta^{ur} = - \frac{1}{2 \kappa^{2}}\,H^{ur} - \frac{1}{2 \kappa^{2}} \int^{r} \nts d r'\,\left( \partial_{i} H^{ui} + \Gamma^{u}_{\nu\lambda}\,H^{\nu\lambda} \right) + \mathcal{O}(r^{0}) ~,
\end{gather}
The part of this shift that diverges as $r\to\infty$ is completely fixed once the $r$-dependence of the fields is specified. But an $r$-independent ambiguity remains; such a term is annihilated by the $\partial_{r}$ in \eqref{eq:GeneralThetauShift}. The choice we make for the finite ($r$-independent) part of $\vartheta^{ur}$ is to retain precisely the contribution from $H^{ur}$. This is motivated by considerations of the variational formulation of the theory \cite{McNees:2025acf}, and is consistent with our expectations for the symplectic pairs appearing in the potential.\,\footnote{This proposal is also consistent with the choice of finite ambiguity made in \cite{Geiller:2021vpg, McNees:2023tus}.} But it is of course possible that a different choice may be needed when considering the integrability of charges.

Implementing this shift, the relevant components of $\widetilde{\Theta}$ are
\begin{gather}\label{eq:ShiftedTheta}
    \widetilde{\Theta}^{u} = o(r^{-1}) 
    \qquad 
    \widetilde{\Theta}^{r} = \Theta^{r} + \partial_{u}\left[\,\frac{1}{2 \kappa^{2}}\,H^{ur} +  \, \frac{1}{2 \kappa^{2}} \int^{r} \nts d r'\,\left( \partial_{i} H^{ui} + \Gamma^{u}_{\nu\lambda}\,H^{\nu\lambda} \right)\right] ~.
\end{gather}
As explained above, the $H^{ur}$ part of the shift in $\widetilde{\Theta}^{r}$ contributes both divergent and finite parts as $r \to \infty$, while the integral captures only terms that diverge in this limit. Note that the shift in the $r$ component of $\widetilde{\Theta}$ is a corner term with support on $\partial \BB = \BB \cap \NN$ where $\NN$ is described on a neighborhood of $\BB$ as a surface of constant $u$.

We now introduce the partial Bondi gauge and describe field configurations satisfying a subset of the equations of motion, before providing explicit results for the shifted symplectic potential \eqref{eq:ShiftedTheta} for these field configurations.

\section{Partial Bondi gauge}
\label{sec:PartialBondiGauge}

In this section we review the partial Bondi gauge \cite{Geiller:2022vto}, along with a set of ``on-shell'' conditions for the fields that are sufficient to implement our prescription for $\widetilde{\Theta}$.

The partial Bondi gauge (PBG) describes solutions of Einstein equations with or without a cosmological constant $\Lambda$. Such spacetimes are asymptotically locally de Sitter (AldS, $\Lambda>0$), anti-de Sitter (AlAdS, $\Lambda<0$), or flat ($\Lambda = 0$). In four dimensions, the line element in partial Bondi gauge is
\begin{gather}\label{eq:PartialBondiMetric}
	ds^{2} = e^{2\,\beta}\,\VV\,du^{2} - 2\,e^{2\beta}\,du\,dr + \gamma_{AB}\,\big(dx^{A} - \UU^{A} du\big)\big(dx^{B} - \UU^{B} du\big) ~,
\end{gather}
where $\VV$, $\UU^{A}$, and $\beta$ are arbitrary function of the coordinates $x^{\mu} = (u,r,x^A)$.

Setting $g_{rr} = g_{rA}=0$ in \eqref{eq:PartialBondiMetric} imposes only three conditions on the fields. This partial gauge fixing is a convenient starting point since it encompasses both Bondi-Sachs (BS) gauge \cite{Bondi:1962px,Sachs:1962zza}, which can be imposed by using the remaining gauge freedom to set $\partial_{r}(r^{-2}\sqrt{\gamma}\,) = 0$, and the generalized Newman-Unti (NU) gauge \cite{Newman:1962cia}, by setting $\partial_{r} g_{ur} = 0$.\,\footnote{Another, more general approach to relaxing the Newman-Unti gauge was investigated in \cite{Campoleoni:2023fug}.} These two gauges are widely used to describe null infinity and have also been used to study Al(A)dS spacetimes \cite{Poole:2018koa, Compere:2019bua, Fiorucci:2020xto, Compere:2020lrt, Compere:2023ktn}. However, relating quantities computed in one gauge to the other can be difficult. A practical advantage of the PBG is that one can perform calculations first and then specialize to either gauge afterwards. One might also avoid complicated transformations need to move between different gauges. For instance the Kerr metric was first brought from Boyer-Lindquist coordinates to PBG \cite{2003CQGra..20.4153F, Venter:2005cs, Hoque:2021nti} before a second transformation was used to reach Bondi-Sachs gauge \cite{Barnich:2011mi}. The ability to work directly in PBG eliminates this last step. Finally, it was shown in \cite{Geiller:2024amx} that additional charges are present in PBG. The physical significance of these charges is still under investigation, but our results in section \ref{sec:SymplecticPotentialInBondiGauge} shed some light on their nature.

Throughout the rest of this paper we enforce the equations of motion conjugate to variables not fixed by \eqref{eq:PartialBondiMetric}, which is sufficient to ensure that the bulk term vanishes in the variation of $L_{M}$: $E^{\mu\nu}\,\delta g_{\mu\nu} = 0$. Such field configurations will be referred to, for lack of a better term, as \textit{partially on-shell}. There are two main reasons for considering this limited set of conditions on the fields, rather than going fully on-shell. First, in lower dimensions there are examples where enforcing a gauge constraint eliminates charges that realize off-shell symmetries of a holographic dual \cite{Grumiller:2017qao}. And second, we would ultimately like to establish a minimal set of conditions for which our prescription leads to a set of integrable charges in absence of radiation.

The bulk term in the variation of the Einstein-Hilbert Lagrangian is
\begin{gather}\label{eq:Emunudeltagmunu}
	E^{\mu\nu}\,\delta g_{\mu\nu} = \frac{1}{2\kappa^{2}}\,\sqrt{-g} \left( \, \frac{1}{2}\,g^{\mu\nu}\big(R - 2\Lambda\big) - R^{\mu\nu}\right) \delta g_{\mu\nu} ~.
\end{gather}
The partially on-shell conditions correspond to enforcing the components $E^{uu}$, $E^{ur}$, $E^{uA}$, and $E^{AB}$ conjugate to variables appearing in \eqref{eq:PartialBondiMetric}, but not the constraints $E^{rr}$ and $E^{rA}$ conjugate to the gauge-fixed components $g_{rr} = g_{rA} = 0$.\,\footnote{In Fefferman-Graham gauge with AdS coordinate $\rho$ and three dimensional coordinates $x^i$, one would enforce $E^{ij} = 0$ but not the $E^{\rho\rho}$ or $E^{\rho i}$ components.}  In partial Bondi gauge these last two equations describe the flux balance laws for the mass and the angular momentum \cite{Geiller:2022vto}.

The focus here is on $\Theta$ and related quantities at $r \to \infty$, so it is sufficient to solve the equations of motion on a neighborhood of $\BB$ at large $r$. To do this we assume a polyhomogeneous expansion of the form~\footnote{This is not the most general starting point. One might consider the appearance of $\ln r$ terms at the first sub-leading order in the large $r$ expansion \cite{Chrusciel:1993hx}, or additional powers of $\ln r$ at $\OO(r^{-3})$ \cite{Kehrberger:2024aak,Geiller:2024ryw}}  
\begin{gather}
    \label{eq:gammaAB}
    \gamma_{AB} = r^{2}\,\left(\gamma^{0}_{AB} + \frac{1}{r}\,\gamma^{1}_{AB} + \frac{1}{r^2}\,\gamma^{2}_{AB} + \frac{1}{r^3}\,\big(\gamma^{3}_{AB} + \AA^{3}_{AB}\,\ln r\big) + o(r^{-3}) \right) ~.
\end{gather}
The tensors $\gamma^{n}_{AB}$ for $n = 1, 2, 3$ are often written in the literature as $C_{AB}$, $D_{AB}$, and $E_{AB}$, respectively. Solving the equations of motion order-by-order in powers of $r$ and $\ln r$, the large-$r$ behaviors of $\beta$, $\UU^{A}$, and $\VV$ take the form
\begin{subequations}\label{expUbetaV}
\begin{align} \label{eq:BetaExpansion}
	\beta = &\,\, \beta_0 + \frac{1}{r}\,\beta_{1} + \frac{1}{r^{2}}\,\beta_2 + \frac{1}{r^{3}}\,\big(\beta_{3} + \tilde{\beta}_{3}\,\ln r \big) + o(r^{-3})  \\ 
    \label{eq:UAExpansion}
 	\UU^{A} = &\,\, \UU_{0}^{A} + \frac{1}{r}\,\UU_{1}^{A} + \frac{1}{r^{2}}\,\UU_{2}^{A} + \frac{1}{r^{3}}\,\big(\UU_{3}^{A} + \tilde{\UU}_{3}^{A}\,\ln r \big) + o(r^{-3}) \\ 
    \label{eq:VExpansion}
	\VV = & \,\, r^{2}\,\left(\VV_0 + \frac{1}{r}\,\VV_{1} + \frac{1}{r^2}\,\VV_{2} + \frac{1}{r^3}\big(\VV_{3} + \tilde{\VV}_{3}\,\ln r\big)+ o(r^{-3})\right) ~.
\end{align}
\end{subequations}
Additional powers of $\ln r$ appear at higher orders in these expansions \cite{Chrusciel:1993hx}, but the quantities we are interested in are completely characterized by the terms shown above. Fall off described as $o(r^{-1})$ in the previous section will generally be $\mathcal{O}(r^{-2})$ or $\mathcal{O}(r^{-2}\ln r)$ for the fields \eqref{eq:gammaAB}-\eqref{expUbetaV}. We now discuss the content of these expansions and the results of enforcing the partially on-shell conditions.

The leading term $\gamma^{0}_{AB}(u,x^{C})$ in \eqref{eq:gammaAB} is the metric on the celestial sphere when $\Lambda = 0$, which is part of a conformal Carrollian structure along with the null congruence defining the asymptotic boundary. When $\Lambda \neq 0$, it is part of a three dimensional metric that is a representative of a conformal class of metrics at conformal infinity. We will refer to $\gamma^{0}_{AB}$ simply as the transverse metric. The two dimensional covariant derivative compatible with the transverse metric is $\DD_{A}$, its curvature is $\RR_{AB}$, and we use $\sqrt{\gamma_0}$ to indicate the square root of its determinant. Throughout this section and the rest of the paper, quantities on the right-hand sides of \eqref{eq:gammaAB}-\eqref{expUbetaV} have two dimensional indices raised and lowered with $\gamma^{0}_{AB}$. This includes traces, which we denote by \begin{gather}
    \gamma_{n} = \gamma_{0}^{AB}\,\gamma^{n}_{AB} \qquad \AA_{3} = \gamma_{0}^{AB}\,\AA^{3}_{AB} ~.
\end{gather}
The trace-free part of a two dimensional tensor is indicated with a hat, as
\begin{gather}
    \hat{\gamma}^{n}_{AB} = \gamma^{n}_{AB} - \frac{1}{2}\,\gamma^{0}_{AB}\,\gamma_{n} \qquad  \hat{\AA}^{3}_{AB} = \AA^{3}_{AB} - \frac{1}{2}\,\gamma^{0}_{AB}\,\AA_{3} ~.
\end{gather}
Brackets $\langle\,\,\rangle$ on a pair of indices refers to the symmetric, trace-free part, defined with an overall factor of $1/2$. A number of useful identities for two dimensional tensors built from the fields $\gamma^{n}_{AB}$, $\beta_{n}$, $\VV_{n}$, $U_{n}^{A}$, and their derivatives, can be found in appendix \ref{app:Identities}.

With the notation described above, $\hat{\gamma}^{1}_{AB}$ is (up to a factor) the shear of the normal to $\BB$. We will frequently make use of the traceless part of $\gamma^{2}_{AB}$ with a term proportional to $\gamma_{1}\,\hat{\gamma}^{1}_{AB}$ removed:
\begin{gather}\label{eq:LABDefinition}
    \LL_{AB} = \hat{\gamma}^{2}_{AB} - \frac{1}{4}\,\gamma_{1}\,\hat{\gamma}^{1}_{AB} ~.
\end{gather}
Although this quantity does not appear with a factor of $\ln r$ in \eqref{eq:gammaAB}, it is sometimes referred to as the ``source'' of log terms, since $\LL_{AB}\neq 0$ generates $\ln r$ terms at orders $r^{-3}$ and higher in the expansions \eqref{expUbetaV}.
In Newman-Penrose formalism \cite{Newman:1961qr, Newman:2009}, $\hat \gamma_{AB}^1$ is related to the spin coefficient $\sigma$, $\VV_3$ to $\text{Re}(\Psi_2)$, $\UU_3^A$ to $\Psi_1$, and $\hat \gamma^3_{AB}$ to $\Psi_0$.

Solving the $E^{u\lambda}$ equations of motion gives algebraic constraints on the coefficients $\beta_n$, $\UU_{n}^{A}$, and $\VV_{n}$ appearing at the first few orders in \eqref{expUbetaV}. They are determined by boundary data $\beta_0$, $\UU_{0}^{A}$, and $\gamma^{0}_{AB}$ on $\BB$, along with quantities that are not constrained by the equations of motion (for instance, $\hat{\gamma}^{1}_{AB}$ when $\Lambda =0$) or the partial gauge fixing. The latter category includes the traces $\gamma_{n}$ ($n \geq 1$), which do not satisfy any conditions in PBG.\,\footnote{The traces may be constrained in a completely fixed gauge such as Bondi-Sachs. Also, in the flat case $\Lambda = 0$ a new gauge fixing was proposed in \cite{Geiller:2024amx} with an enhancement of symmetries with respect to BMS symmetries. This new gauge fixes the traces $\gamma_{AB}^n$ with $n>2$, leaving $\gamma_1$ and $\gamma_2$ arbitrary.} Solving $E^{AB}$ constrains the evolution of the terms appearing in \eqref{eq:gammaAB}. Dynamical quantities appear at order $r^{-3}$ in the expansions \eqref{expUbetaV} for $\UU^{A}$, $\VV$, and $\gamma_{AB}$.~\footnote{The terminology here implies a choice about which  quantities are being held fixed on $\BB$. For convenience, we refer to the leading terms in \eqref{expUbetaV} as boundary data, and describe integration constants at subleading order as appearing dynamically. The latter will have their evolution in time constrained by the $E^{rr}$ and $E^{rA}$ equations of motion. One might instead consider boundary conditions which reverse (or mix) these roles.}  Subsequent terms in the expansions depend on these quantities, as well. We will only comment here on some key results at the first few orders in the large-$r$ expansion of the equations of motion, but detailed results for all quantities in \eqref{expUbetaV} can be found in appendix \ref{app:EOMPBG}.

First, the equations $E^{uu}$ and $E^{uA}$ are independent of the cosmological constant $\Lambda$, so we can use their results in all the various settings we consider. They fix coefficients of sub-leading terms in the large-$r$ expansions of $\beta$ and $\UU^{A}$, respectively. The equation $E^{uu}$ for $\beta$ is first order in $\partial_r$, with integration constant $\beta_0$ appearing as boundary data at leading order in \eqref{expUbetaV}. The first subleading term in the expansion of $\beta$ vanishes,
\begin{gather}
    \beta_1 = 0 ~,
\end{gather}
and subsequent terms $\beta_{n}$ ($n \geq 2$) are determined algebraically by the $\gamma^{n}_{AB}$. For example,
\begin{gather}
    \beta_{2} = \frac{1}{32}\,\hat{\gamma}_{1}^{AB}\,\hat{\gamma}^{1}_{AB} + \frac{1}{64}\,\big(\gamma_{1})^{2} - \frac{1}{8}\,\gamma_2 ~.
\end{gather}
The equation $E^{uA}$ for $\UU^{A}$ is second order in $\partial_r$. One of the two integration constants appears as the boundary data $\UU_{0}^{A}$ at leading order. The first few subleading terms in $\UU^{A}$ are determined by $\gamma^{0}_{AB}$, $\beta_{0}$, and the shear $\hat{\gamma}^{1}_{AB}$, and there is a log term at order $r^{-3}$ which is sourced by the (two-dimensional) divergence $\DD^{B}\LL_{AB}$. Complete expressions for these quantities can be found in appendix \ref{app:EOMPBG}. The second integration constant, $\UU_{3}^{A}$, appears dynamically at order $r^{-3}$ and encodes information about the angular momentum.

The $E^{ur}$ equation of motion is solved for $\VV$ and fixes the first three coefficients in \eqref{eq:VExpansion}. This equation \textit{does} depend on the cosmological constant, and yields qualitatively different results for $\Lambda =0$ and $\Lambda \neq 0$. The leading order term in $E^{ur}=0$ gives
\begin{align}\label{eq:V0Eqn}
\VV_{0} = & \,\, \frac{\Lambda}{3}\,e^{2\beta_0} ~.
\end{align}
As a result, the $r \to \infty$ limit of the ``unphysical'' line element $r^{-2}\,ds^{2}$ on a surface of constant $r$ is degenerate for $\Lambda = 0$
\begin{equation}
\lim_{r\to\infty} \frac{ds^2}{r^2}=e^{2\beta_0}\VV_0 \,du^2+\gamma^0_{AB}\big(dx^A-\UU_0^Adu\big)\big(dx^B-\UU_0^Bdu\big) ~.
\end{equation}
When $\Lambda \neq 0$ it is instead a representative of a conformal class of metrics at conformal infinity. As an equation for $\VV$, $E^{ur}$ is linear in $\partial_{r}$. The integration constant $\VV_{3}$ appears dynamically at order $r^{-3}$ and encodes information about the mass of the spacetime.

The $E^{AB}$ equation of motion can be decomposed into trace and trace-free parts with respect to $g^{AB} = \gamma^{AB}$. The trace depends on $\Lambda$, but the trace-free part does not. However, the trace-free equation gives conditions which depend on the $\VV_{n}$, which in turn depend on the cosmological constant. Once the $E^{ur}$ equation and the trace-free part of $E^{AB}$ are enforced, the trace part of $E^{AB}$ is automatically satisfied. After imposing the $E^{uu}$ and $E^{uA}$ equations, the trace-free part of $E^{AB}$ reduces to 
\begin{gather}\label{eq:ETT}
    R_{AB} - \frac{1}{2}\,\gamma_{AB}\,\gamma^{CD}\,R_{CD} = 0 ~,
\end{gather}
where $R_{AB}$ are components of the four dimensional Ricci tensor. The leading part of this equation is
\begin{gather}\label{eq:LeadingEAB}
    \partial_{u}\gamma^{0}_{AB} - \gamma^{0}_{AB}\,\partial_{u}\ln\sqrt{\gamma_0} + 2\,\DD_{\langle A} \UU^{0}_{B\rangle} = \VV_{0}\,\hat{\gamma}^{1}_{AB} ~.
\end{gather}
The implications of this equation depend on the value of $\Lambda$. Specifically, the right-hand side is proportional to $\Lambda$ and multiplies the shear $\hat\gamma^1_{AB}$. Therefore, for $\Lambda \neq 0$, the shear is determined by the left-hand side of the equation, consistent with the observation that this combination of the boundary data should be non-vanishing for radiation in Al(A)dS spacetimes \cite{Compere:2019bua, Compere:2020lrt,Bonga:2023eml,Ciambelli:2024kre, Fernandez-Alvarez:2021yog}. For $\Lambda = 0$, the equation implies that the shear remains unconstrained and that the left-hand side must vanish. In this case, the left-hand side can be interpreted as the shear of the null congruence defining the asymptotic boundary.

At higher orders, we obtain conditions like
\begin{align}\label{logeom}
    0 &= \VV_{0}\,\LL_{AB}\,,\qquad 0 =\partial_{u} \LL_{AB} + \pounds_{\UU_0}\LL_{AB} + \VV_{0}\,\hat{\AA}^{3}_{AB} ~.
\end{align}
For non-vanishing cosmological constant, with $\VV_0 \neq 0$, this implies that $\LL_{AB}$ and $\hat{\AA}^{3}_{AB}$ both vanish. However, when $\Lambda = 0$, these quantities are generically present \cite{Chrusciel:1993hx,Godazgar:2020peu,ValienteKroon:2001pc, Geiller:2022vto, Kehrberger:2024aak, Geiller:2024ryw} and satisfy first order differential equations involving their derivatives in the $u$ and $\UU_{0}^{A}$ directions. For $\Lambda = 0$, subsequent orders of this equation give the evolution of $\gamma^n_{AB}$ with $n \geq 3$ \cite{Freidel:2021ytz,Geiller:2024bgf,Kmec:2024nmu}. For non-vanishing cosmological constant, the function $\gamma^3_{AB}$ is unconstrained and the higher order terms in \eqref{eq:gammaAB} are fixed algebraically in terms of quantities (and their derivatives) with $n \leq 3$ \cite{Compere:2019bua}.

Terms in \eqref{eq:gammaAB}-\eqref{expUbetaV} relevant for the $r \to \infty$ limit of the symplectic potential appear at orders $n \leq 3$. After imposing the partially on-shell conditions, these are the boundary sources $\beta_0$, $\UU_0^A$, and the transverse metric $\gamma_{AB}^0$; the shear $\hat \gamma_{AB}^1$; the traces $\gamma_1$ and $\gamma_2$; the source of log terms $\LL_{AB}$; $\hat \gamma_{AB}^3$ through the spin-2 charge; $\VV_{3}$, $\gamma_3$, and $\AA_{3}$ through the mass; and $\UU_3^A$ through the angular momentum.

\section{Symplectic potential in partial Bondi gauge}
\label{sec:SymplecticPotentialInBondiGauge}
In this section we compute the bare symplectic potential in partial Bondi gauge, including all contributions which diverge or remain finite as $r \to \infty$, then apply the prescription of section \ref{sec:PrescriptionForEH} to obtain the shifted potential $\widetilde{\Theta}$. All divergent terms in $\widetilde{\Theta}$ vanish for $\Lambda = 0$, or reduce to $\delta$-exact quantities when $\Lambda \neq 0$. We then discuss the different symplectic pairs appearing in the potential. Note that we allow arbitrary time dependence for the boundary sources, consistent with \eqref{eq:LeadingEAB}. Throughout this section we set $\kappa^{2} = 8\pi G = 1$.

\subsection{Bare symplectic potential}
In partial Bondi gauge \eqref{eq:PartialBondiMetric}, the components of $H^{\mu\nu}$ entering in the definition \eqref{def:symplpot} of the bare symplectic potential are
\begin{align}\label{eq:Hur}
    H^{ur} = &\,\, 2\,\sqrt{\gamma}\,\delta\beta + 2\,\delta \sqrt{\gamma}\\ \label{eq:Hrr}    
    H^{rr} = &\,\, \delta\Big(\sqrt{\gamma}\,\VV \Big) + 2\,\sqrt{\gamma}\,\VV\,\delta\beta + \VV\,\delta\sqrt{\gamma} \\ \label{eq:HrA}
    H^{rA} = &\,\, \delta\Big(\sqrt{\gamma}\,\UU^{A} \Big) + 2\,\sqrt{\gamma}\,\UU^{A}\,\delta\beta + \UU^{A}\,\delta\sqrt{\gamma} \\ \label{eq:HAB} 
    H^{AB} = &\,\, e^{2\beta} \, \sqrt{\gamma} \, \Big( \big( \gamma^{AC} \gamma^{BD} - \gamma^{AB} \gamma^{CD} \big) \delta \gamma_{CD} - 4\,\gamma^{AB} \, \delta\beta \Big)
\end{align}
The $H^{uu}$ and $H^{uA}$ components both vanish since $g^{uu} = g^{uA} = 0$.

The $\vartheta^{ur}$ component of the shift two-form is extracted from the $u$ component of the bare symplectic potential. Several components of $H^{u\nu}$ and $\Gamma^{u}_{\nu\lambda}$ vanish, leaving just
\begin{align}
	\Theta^{u} &= \frac{1}{2}\,\partial_{r} H^{ur} + \frac{1}{2}\,\Gamma^{u}_{AB}\,H^{AB} ~\\
& = \partial_{r} \Big(  \sqrt{\gamma}\,\delta \beta + \delta\sqrt{\gamma} \Big) + \sqrt{\gamma}\left(\frac{1}{4} \partial_{r} \gamma_{AB}\,\big(\gamma^{AC}\gamma^{BD} - \gamma^{AB}\gamma^{CD}\big)\,\delta \gamma_{CD} - 2\,\partial_{r}\ln\sqrt{\gamma}\,\delta\beta \right) \,.
\end{align}
Contributions to $\vartheta^{ur}$ from the second set of terms, which are not already a total $r$-derivative, are captured by the integral in \eqref{eq:varthetapresc}. Only terms with fall off slower than $o(r^{-1})$ are relevant. Using the expansions \eqref{eq:gammaAB}-\eqref{expUbetaV} and partially on-shell conditions, we have
\begin{align}\label{Thetau}
	\Theta^{u} = & \,\, \partial_{r}\Big(  \sqrt{\gamma}\,\delta \beta+\delta\sqrt{\gamma} \Big) \\ \nonumber
 & + r\,\Big( - \delta\sqrt{\gamma_0} - 4\,\sqrt{\gamma_0}\,\delta\beta_0 \,\Big) - \frac{1}{4} \, \sqrt{\gamma_0} \, \hat{\gamma}_{1}^{AB} \delta\gamma^{0}_{AB} - \frac{1}{4}\,\gamma_{1}\delta\sqrt{\gamma_0} - \frac{1}{2}\,\sqrt{\gamma_0}\,\delta\gamma_1 - \sqrt{\gamma_0}\,\gamma_1\,\delta\beta_0 \\ \nonumber
 & + \frac1{r} \sqrt{\gamma_0}\,\left(- \frac{1}{2}\,\LL^{AB} \delta\gamma_{AB}^0 \right) +o(r^{-1}) ~.
\end{align}
The last two lines contribute to $\vartheta^{ur}$ at orders $r^2$, $r^1$, and $\ln r$.

The $r$ component of the bare symplectic potential is more involved. Only a few terms vanish in partial Bondi gauge, leaving
\begin{align}
    \Theta^{r} = &\,\, \frac{1}{2}\,\partial_{r} H^{rr} + \frac{1}{2}\,\partial_{u} H^{ru} + \frac{1}{2}\,\partial_{A} H^{rA} + \frac{1}{2}\,\Gamma^{r}_{rr} H^{rr} + \Gamma^{r}_{ru} H^{ru} + \Gamma^{r}_{rA} H^{rA} +\frac{1}{2}\,\Gamma^{r}_{AB} H^{AB} \\ \nonumber
    = & \,\,  \partial_{u}\Big(\sqrt{\gamma}\,\delta\beta + \delta \sqrt{\gamma}\Big)  + \partial_{A} \left( \frac{1}{2}\,\delta\big(\sqrt{\gamma}\,\UU^{A} \big) + \sqrt{\gamma}\,\UU^{A}\,\delta\beta + \frac{1}{2}\,\UU^{A}\,\delta\sqrt{\gamma} \right) \\ \nonumber
    &\,\, + \frac{1}{2}\,\partial_{r} H^{rr} + \frac{1}{2}\,\Gamma^{r}_{rr} H^{rr} + \Gamma^{r}_{ru} H^{ru} + \Gamma^{r}_{rA} H^{rA} + \frac{1}{2}\,\Gamma^{r}_{AB} H^{AB}
\end{align}
Using the large-$r$ expansions for the fields leads to an expression of the form
\begin{align}\label{eq:PreThetar}
    \Theta^{r} = &\,\, \Theta^{r}_{0} + \partial_{A} B^{A} + \ln r\,\sqrt{\gamma_0}\,\DD^{B}\LL_{AB}\,\delta\UU_{0}^{B} \\ \nonumber
   &\,\, +  \delta\Big(r^{3}\,l_{\BB}^{\,3} + r^{2}\,l_{\BB}^{\,2} + r \,l_{\BB}^{\,1} + \ln r\,l_{\BB}^{\ln}\Big)  + \partial_{u} \Big( r^{2}\,f_{\partial \BB}^{\,2} + r\,f_{\partial \BB}^{\,1}\Big) + o(r^0) ~.
\end{align}
Here $\Theta^{r}_{0}$ represents all finite $\OO(r^0)$ terms in $\Theta^{r}$, including both terms on $\BB$ and corner terms with support on $\partial \BB$. We postpone writing the full expression for $\Theta^{r}_{0}$, which is complicated, until section \ref{sec:finitesympl}. There is also a total derivative on the sphere, $\partial_{A} B^{A}$, with
\begin{align}\nonumber
    B^A & = \frac{1}{2}\,\sqrt{\gamma} \, \delta \UU^{A} + \UU^{A}\,\delta\sqrt{\gamma} + \sqrt{\gamma} \, \UU^{A} \, \delta\beta \\
        & + r \left[ \,\sqrt{\gamma_0} \left( \frac{1}{2}\,\hat\gamma_{1}^{A}{}_{B} \, \delta\UU_{0}^{B} + \frac{1}{4} \, \UU^A_0 \hat\gamma_1^{BC} \delta \gamma_{BC}^0 \right) - \frac{1}{4} \gamma_1 \delta\big( \sqrt{\gamma_0} \, \UU^A_0\big) \right. \\ \nonumber
        & \qquad +\left.\sqrt{\gamma_0}\left(\frac12(\delta\gamma_0)^{AB}\DD_{B}e^{2\beta_0}-\frac12e^{2\beta_0}\DD_{B}(\delta\gamma_0)^{AB} + e^{2\beta_0}\DD^{A}\delta\ln\sqrt{\gamma_0} - \delta\ln\sqrt{\gamma_0}\DD^{A}e^{2\beta_0}\right)  \right] ~.
\end{align}
Such terms vanish when $\Theta^{r}$ is integrated over $\BB$, so we discard them henceforth. The remaining terms in \eqref{eq:PreThetar} all diverge at $r\to\infty$. The coefficients of the $\delta$-exact terms in the second line of \eqref{eq:PreThetar} are given by 
\begin{align}
    l_{\BB}^{\,3} = & \,\, \sqrt{\gamma_0}\,\VV_{0} \\
    l_{\BB}^{\,2} = & \,\, \frac{3}{2}\,\sqrt{\gamma_0}\,\VV_{1} \\    
    l_{\BB}^{\,1} = & \,\, \sqrt{\gamma_0}\,\left[\VV_{2} + \frac{1}{2}\,\gamma_{1}\,\VV_{1} + \frac{1}{2}\,e^{2\beta_0}\,\RR + \frac{1}{4}\,\VV_{0} \, \left(\hat{\gamma}_{1}^{AB} \hat{\gamma}^{1}_{AB} - \frac{1}{2} (\gamma_1)^{2} \right) \right. \\ \nonumber
    & \,\, \qquad \quad \left. + \frac{1}{4}\,\gamma_{1}\,\Big( \partial_{u}\ln\sqrt{\gamma_0} + \DD_{A} \UU_{0}^{A} \Big) - \frac{1}{4}\,\hat{\gamma}_{1}^{AB}\,\Big(\partial_{u}\gamma^{0}_{AB} + \DD_{A} \UU^{0}_{B} + \DD_{B}\UU^{0}_{A}\Big) \right]\\
    l_{\BB}^{\ln} = &\,\, \frac{\Lambda}{4}\,\sqrt{\gamma_0}\,e^{2\beta_0}\,\AA_{3} ~.
\end{align}
In the variational problem, these contributions to the first variation of the action must be addressed with an appropriate set of surface terms on $\BB$. Since they are $\delta$-exact, they do not contribute to the symplectic current $\omega$ or codimension-2 form $k$. On the other hand, the corner terms in the third line of \eqref{eq:PreThetar} are \textit{not} $\delta$-exact. They are given by
\begin{align}
    f_{\partial\BB}^{\,2} = & \,\, \delta\sqrt{\gamma_0} + \sqrt{\gamma_0}\,\delta\beta_0 \\
    f_{\partial\BB}^{1} = & \,\, \frac{1}{4}\,\sqrt{\gamma_0}\,\hat{\gamma}_{1}^{AB}\,\delta\gamma^{0}_{AB} - \frac{1}{4}\,\gamma_{1}\,\delta\sqrt{\gamma_0} + \frac{1}{2}\,\delta\Big(\sqrt{\gamma_0}\,\gamma_{1} \Big) + \frac{1}{2}\,\sqrt{\gamma_0}\,\gamma_{1}\,\delta\beta_0 ~.
\end{align}
These terms will be corrected in $\widetilde{\Theta}^{r}$ by contributions from the shift $\vartheta^{ru}$. This leaves the last term in the first line of \eqref{eq:PreThetar} -- a $\ln r$ divergence related to the source of log terms $\LL_{AB}$. Though it is not obvious, this is also a corner term. Using the partially on-shell conditions for the fields and their variations, the result \eqref{eq:LogIdentity} allows this term to be rewritten. Then the structure of $\Theta^{r}$ is
\begin{align}\label{eq:Thetar}
    \Theta^{r} \oeq &\,\, \Theta^{r}_{0} + \delta l_{\BB} + \partial_{u} \left( r^{2}\,f_{\partial \BB}^{\,2} + r\,f_{\partial \BB}^{\,1} + \ln r \, \frac{1}{2}\,\sqrt{\gamma_0}\,\LL^{AB} \delta\gamma^{0}_{AB} \right) + o(r^0) ~.
\end{align}
where $\oeq$ indicates equality up to total derivatives on the sphere, and $\delta l_{\BB}$ is the full set of divergent $\delta$-exact terms appearing in \eqref{eq:PreThetar}.

\subsection{Applying the prescription}
The $\vartheta^{ur}$ component of the shift two-form is extracted from $\Theta^{u}$ via the prescription \eqref{eq:varthetapresc}. The term in the first line of \eqref{Thetau}, which comes from $\partial_{r}H^{ur}$, contributes both divergent and finite parts, while the terms in the second and third lines are captured by the integral in \eqref{eq:varthetapresc} and contribute only divergent parts. The result is
\begin{align}\label{eq:varthetaur}
   \vartheta^{ur} = &\,\, r^{2}\,\left(-\frac{1}{2}\,\delta\sqrt{\gamma_0} + \sqrt{\gamma_0}\,\delta\beta_0\right) \\
   &\,\, + r\,\left(\frac{1}{4}\,\sqrt{\gamma_0} \,\hat{\gamma}_{1}^{AB} \delta\gamma^{0}_{AB} - \frac{1}{4}\,\gamma_1\,\delta\sqrt{\gamma_0} +  \frac{1}{2}\,\sqrt{\gamma_0}\,\gamma_{1}\,\delta\beta_0 \right) \\
   &\,\, + \ln r \,\left(\frac{1}{2}\,\sqrt{\gamma_0}\,\LL^{AB}\,\delta\gamma^{0}_{AB}\right) \\
   & \,\, + \sqrt{\gamma_0}\,\left( -\frac{3}{8}\,\delta\gamma_2 - \frac{1}{2}\,\gamma_{2}\,\delta\ln\sqrt{\gamma_0} - \frac{1}{2}\,\gamma_2\,\delta\beta_0 - \frac{1}{64}\,\delta\big(\gamma_{1}^{\,2}\big) + \frac{7}{16}\,\hat{\gamma}_{1}^{AB}\,\delta\hat{\gamma}^{1}_{AB} \right. \\ \nonumber
   & \qquad \qquad \quad \left. - \frac{3}{16}\,\delta\ln\sqrt{\gamma_0}\, \hat{\gamma}_{1}^{AB}\,\hat{\gamma}^{1}_{AB} + \frac{1}{4}\,\hat{\gamma}_{1}^{AB}\,\hat{\gamma}^{1}_{AB}\,\delta\beta_{0}\right) ~. 
\end{align}
Then $\widetilde{\Theta}^{u} = \Theta^{u} + \partial_{r}\vartheta^{ur} = o(r^{-1})$.

After making the shift to $\widetilde{\Theta}$, the $r$ component is $\widetilde{\Theta}^{r} = \Theta^{r} + \partial_{u}\vartheta^{ru}$ with $\vartheta^{ru} = - \vartheta^{ur}$. This corrects the corner terms in $\Theta^{r}$, giving
\begin{gather}
 \widetilde{\Theta}^{r} \oeq \widetilde{\Theta}_{0}^{r} + \delta\,\Big( l_{\BB} + \partial_{u}  l_{\partial\BB} \Big) + o(r^0) ~.
\end{gather}
The corner terms, which are now $\delta$-exact, are 
\begin{gather}
    l_{\partial\BB} = r^{2}\,\frac{3}{2}\,\sqrt{\gamma_0} + r\,\frac{1}{2}\,\sqrt{\gamma_0}\,\gamma_1  ~,
\end{gather}
and the $\ln r$ term in \eqref{eq:Thetar} has cancelled. The finite part has also been shifted, and is related to $\Theta_{0}^{r}$ by
\begin{gather}
    \widetilde{\Theta}_{0}^{r} = \Theta_{0}^{r} - \partial_{u} \vartheta_{0}^{ur} ~,
\end{gather}
where $\vartheta_{0}^{ur}$ indicates the $\OO(r^{0})$ terms in \eqref{eq:varthetaur}.

Since all potential divergences are now $\delta$-exact, this result is sufficient to ensure finite $\omega$ and $k$ at $r \to \infty$. However, the cancellations are more extensive than just the $\ln r$ corner term. Using the partially on-shell conditions \eqref{eq:V1eqn} for $\VV_{1}$ and \eqref{eq:V2eqn} for $\VV_{2}$, we find
\begin{gather}
    l_{\BB} = \tilde{l}_{\BB} + \partial_{u}\big(-l_{\partial\BB}\big) ~.
\end{gather}
That is, the divergent corner terms cancel entirely, leaving
\begin{gather}
 \widetilde{\Theta}^{r} \oeq \widetilde{\Theta}_{0}^{r} + \delta \,\tilde{l}_{\BB} + o(r^0) ~,
\end{gather}
with $\tilde{l}_{\BB}$ now given by
\begin{equation}
 \tilde{l}_{\BB}= \frac{\Lambda}{4}  e^{2\beta_0} \sqrt{\gamma_0}\,\left[ \,\frac43\,r^3+\gamma_{1}\,r^2+\left(\gamma_{2} + \frac{1}{8}\,(\gamma_1)^{2} - \frac{3}{4}\,\hat{\gamma}_{1}^{AB} \, \hat{\gamma}^{1}_{AB} \right)\,r+\AA_{3}\,\ln r \right]
\end{equation}
An immediate consequence is that the shifted potential $\widetilde{\Theta}$ is finite as $r \to \infty$ when $\Lambda = 0$. The divergent terms are not just $\delta$-exact in that case, but precisely zero. When $\Lambda \neq 0$ the divergent corner terms vanish, but there are still $\delta$-exact divergent terms on $\BB$. Such terms, which have no effect on $\omega$ and $k$, are relevant for the variational formulation of the theory and may be eliminated by adding an appropriate set of surface terms to the action \cite{McNees:2025acf}.

\subsection{The finite part of the symplectic potential}
\label{sec:finitesympl}

Organizing the terms in the finite part of the symplectic potential is tedious, and the intermediate steps are not illuminating. A \textsc{Mathematica} notebook with the complete expression for $\widetilde{\Theta}^{\,r}_{0}$ is included with the ar$\chi$iv upload as an ancillary file.

Our primary goal in this section is to demonstrate the physical content of $\widetilde{\Theta}_{0}^{r}$ by writing it as a sum of symplectic pairs $p^{i}\,\delta q_{i}$ up to $\delta$-exact terms. From previous work in Newman-Unti and Bondi-Sachs gauges, we know the $q_i$ should include the transverse metric $\gamma^{0}_{AB}$, volume element $\sqrt{\gamma_0}$, and shear $\hat{\gamma}^{1}_{AB}$. However, the role played by the boundary sources $\beta_{0}$ and $\UU_{0}^{A}$ has not been established\,\footnote{However, see recent work in \cite{Riello:2024uvs}.}, and previous results for asymptotically flat spacetimes place conditions on the transverse metric. The result below includes contributions from the full set of boundary sources and allows general variations of an unconstrained transverse metric. A secondary goal is to provide a compact expression for $\widetilde{\Theta}^{r}_{0}$ that applies in both the $\Lambda = 0$ and $\Lambda \neq 0$ cases. This allows us to investigate whether the $\Lambda \to 0$ limit recovers the full $\Lambda = 0$ result. We find this is not the case.

To achieve these goals it is sufficient to focus on the case $\UU_{0}^{A}=0$. This choice, which is consistent with the partially on-shell conditions and requires no additional restrictions on the boundary data, simplifies the organization of the finite terms into symplectic pairs and $\delta$-exact parts. But the condition is sufficiently mild that we can still describe the changes to the potential for non-vanishing $\UU_{0}^{A}$. We emphasize that results and expressions in the previous sections do not make this assumption, and apply for completely general boundary data.

The $p^{i}$ in the symplectic pairs are most conveniently expressed in terms of quantities appearing the Newman-Penrose scalars. First we introduce the so-called covariant mass $\MM$ and spin 2 charge $\EE_{AB}$, related to the Newman-Penrose scalars $\text{Re}\,\Psi_2$ and $\Psi_0$, respectively\,\footnote{The covariant mass is the real coefficient of $\Psi_2$ at order $1/r^3$, while the spin 2 charge appears as $\Psi_0= \EE_{AB} m^Am^B\,r^{-5}+o(r^{-5})$ where $m^A$ is the complex dyad of the Bondi null tetrad.
},
\begin{align}
 \MM &=\frac{1}{2} \, e^{-2\beta_0} \,\VV_{3} + \frac{1}{4}\,e^{-2\beta_0} \frac{1}{\sqrt{\gamma_0}}\,\partial_{u}\left[\sqrt{\gamma_0}\left(\gamma_2 - \frac{1}{8}\,(\gamma_1)^{2} - \frac{1}{4}\,\hat{\gamma}_{1}^{CD}\,\hat{\gamma}^{1}_{CD} \right)\right]\\\nonumber & -\DD_B\hat\gamma^{AB}_1\partial_A\beta_0+\frac12 \partial_A\gamma_1\partial^A\beta_0
 +\frac{\Lambda}{24} \left( \AA_3-4  \gamma_{3} \right) \\ 
 \EE_{AB} & = 3 \, \hat\gamma^3_{AB}-\frac52 \,\hat \AA^3_{AB}-\frac12 \,\gamma_1\,\LL_{AB}-\frac3{4}\,\hat\gamma^1_{AB}\left(\gamma_2 - \frac{1}{8}\,(\gamma_1)^{2} - \frac{1}{4}\,\hat{\gamma}_{1}^{CD}\,\hat{\gamma}^{1}_{CD} \right)
 \end{align}
Note that $\EE_{AB}$ is always accompanied by a factor of $\Lambda$ in $\widetilde{\Theta}_{0}^{r}$, and the quantities $\Lambda \, \hat{\AA}^{3}_{AB}$ and $\Lambda\,\LL_{AB}$ vanish by the equations of motion. Next we consider the news, which is related to the time derivative of the shear. There are various conventions for the news tensor in the literature. Here, we use 
\begin{equation}
       \hat{N}_{AB} = \partial_{u}\hat{\gamma}^{1}_{AB} - \frac{1}{2}\,\hat{\gamma}^{1}_{AB}\,\partial_{u}\ln\sqrt{\gamma_0} - \frac{\Lambda}{6} \,\gamma^{0}_{AB} \, e^{2\beta_0} \hat{\gamma}_{1}^{CD} \, \hat{\gamma}^{1}_{CD} ~,
\end{equation}
which is traceless when $\UU_{0}^{\,A} = 0$.

To ensure $\widetilde{\Theta}$ has an unambiguous $\Lambda \to 0$ limit,\footnote{Preserving the $\Lambda \to 0$ limit for both the action and its first variation requires a careful analysis, and may restrict the space of field configurations. This will be discussed in \cite{McNees:2025acf}.} the equation of motion \eqref{eq:ETT} is applied in a manner consistent with either $\Lambda = 0$ or $\Lambda \neq 0$. For instance, occurrences of $\VV_{0} \, \hat{\gamma}^{1}_{AB}$ are replaced with the right hand side of \eqref{eq:LeadingEAB}, but terms $\hat \gamma_{AB}^1$ without a factor of $\VV_0$ (or $\Lambda$) are not. In this way we avoid dividing by $\Lambda$ and generating terms that are ill-defined when $\Lambda \to 0$. Along the same lines, terms proportional to $\VV_0\,\LL_{AB}$ and $\VV_0\,\hat{\AA}^{3}_{AB}$ are eliminated (since they vanish either by $\Lambda = 0$, or else by \eqref{eq:LABwithCC}-\eqref{eq:hatA3ABwithCC} when $\Lambda \neq 0$) but terms without the factor of $\VV_0$ are left untouched. These same conditions may also be applied to derivatives or variations of the fields.

The finite part of the symplectic potential $\widetilde{\Theta}_{0}^{r}$ for $\UU_0^A=0$ is then\,\footnote{For $\Lambda =0$ the metric on $\BB$ is degenerate. In that case, the appropriate volume element is given in terms of determinant of the transverse metric $\gamma^0_{ab}$ and of the normal volume element $e^{2\beta_0}$ by $\sqrt{\gamma_0} \, e^{2\beta_0}$, see for instance \cite{Hopfmuller:2016scf}. For $\Lambda \neq 0$ the volume element on $\BB$ in PBG is proportional to $\sqrt{\gamma_0}\,e^{2\beta_0}$.}
\begin{empheq}[box=\fbox]{align}
\label{eq:finitesymplpot}
 \raisebox{2em}{~}\quad \widetilde{\Theta}_{0}^{r} & \oeq \frac{1}{4} \sqrt{\gamma_0} \, e^{2\beta_0} \left[ \vphantom{\frac{1}{4}} \, \hat{N}^{AB} \, \delta \left( e^{-2\beta_0} \hat\gamma_{AB}^1 \right) +
       \hat T^{AB} \delta\gamma_{AB}^0 +T \, \delta \big(\ln\sqrt{\gamma_0}-4\beta_0 \big) \,\right] \quad  \\
   \nonumber    
   & + \partial_{u}\left[\, \frac{1}{4}\,\sqrt{\gamma_0} \Big( \hat \tau^{AB}\delta \gamma_{AB}^0+ \tau \,\delta \big(\ln\sqrt{\gamma_0}-4\beta_0 \big)\Big)\right] + \delta \big(A_{\BB} + \partial_u A_{\partial\BB} \big) \raisebox{-1.5em}{~}~.
\end{empheq}
 On $\BB$ we have
\begin{align} 
\label{eq:ThatAB}
\hat T_{AB} = & \,\, e^{-2\beta_0} \partial_{u}\ln\sqrt{\gamma_0} \, \LL_{AB} + \RR \,\hat{\gamma}^{1}_{AB} -\frac{1}{2}\,\DD^{2}\hat{\gamma}^{1}_{AB}+ \frac{\Lambda}{12}\,\hat{\gamma}^{1}_{AB}\,  \hat{\gamma}_{1}^{CD}\,\hat{\gamma}^{1}_{CD}     - \frac{\Lambda}{3} \,\EE_{AB}
\\ \nonumber & + \frac{3}{2}\, e^{-2\beta_0}\,\hat{\gamma}^{1}_{AB}\,\DD^2e^{2\beta_0} -6 \, \partial_C\beta_0 \, \DD^C\hat\gamma_{AB}^1
+8 \, \partial^C\beta_0 \, \DD_{\langle A}\hat\gamma^1_{B\rangle C} \\
\label{eq:Ttrace}
T = & \,\, 4\,\MM - \frac{e^{-2\beta_0}}{2\sqrt{\gamma_0}}\,\partial_{u}\Big(\sqrt{\gamma_0}\,\hat{\gamma}_{1}^{CD}\,\hat{\gamma}^{1}_{CD} \Big) +\DD_{A}\DD_{B} \hat{\gamma}_{1}^{AB}-e^{-2\beta_0}\hat\gamma_1^{AB}\DD_A\partial_{B}e^{2\beta_0} ~,
\end{align}
while the terms with support on the corner $\partial \BB$ involve
\begin{equation}
\hat \tau_{AB}=\frac1{4}\,\gamma_1\,\hat\gamma_1^{AB}\,,\qquad \tau = -\gamma_2+\frac12 \hat{\gamma}_{1}^{AB} \hat{\gamma}^{1}_{AB} ~.
 \end{equation}
The finite part of the symplectic potential includes $\delta$-exact terms with support on $\BB$ and $\partial \BB$. They are given by
\begin{align}
    A_{\BB} = &\,\, \sqrt{\gamma_0}\,e^{2\beta_0}\left[ \,\MM + \frac{\Lambda}{16}\,\gamma_{1} \left( \gamma_2 - \frac{3}{4}\,\hat{\gamma}_{1}^{AB}\,\hat{\gamma}^{1}_{AB} - \frac{1}{24}\,(\gamma_{1})^{2} \right) + \frac{\Lambda}{4}\,\gamma_{3} \,\right] \\
    A_{\partial \BB} = &\,\,\sqrt{\gamma_0}\left[ \, \frac{1}{2}\,\gamma_2 - \frac{3}{8} \, \hat{\gamma}_{1}^{AB}\,\hat{\gamma}^{1}_{AB} \right]
\end{align}
Besides introducing some new terms in the expressions for the $p^{i}$, the effect of $\UU_0^A$ is to add another symplectic pair on $\BB$ of the form
     \begin{align}
\sqrt{\gamma_0} \, \PP_A \, \delta \UU^A_0 ~.
 \end{align}
Here $\PP_A$ is the covariant angular momentum, see for example Eq. (2.40) of \cite{Geiller:2022vto}.

Note that the pairs $p^{i}\,\delta q_{i}$ are only defined up to $\delta$-exact terms at the level of the potential, and may be shifted by moving quantities in and out of $A_{\BB}$ and $A_{\partial\BB}$. This is relevant for the variational formulation of the theory (and is necessary for recovering standard AlAdS results when $\Lambda \neq 0$) but does not affect the current $\omega$ or codimension-2 form $k$. To discuss the symplectic pair content, we need to address the flat $\Lambda =0$ and $\Lambda \neq 0$ cases separately. We start with the flat ($\Lambda = 0$) case. 

\subsubsection*{Symplectic pairs for $\Lambda=0$}
Ignoring the $\delta$-exact terms, the structure of the symplectic potential is 
\begin{equation}
  \widetilde{\Theta}_{0}^{r} \oeq \sqrt{\gamma_0}\,e^{2\beta_0} \,p_i \, \delta q^i + \partial_u\big( \sqrt{\gamma_0}\, p_i^c \, \delta q_c^i \,\big) 
\end{equation}
On $\BB$ one has (up to numerical factors)
\begin{align}
& q^i=\big\{ e^{-2\beta_0} \hat \gamma_{AB}^1, \,\gamma_{AB}^0, \, \ln\sqrt{\gamma_0}-4\beta_0 , \,\UU_0^A \big\} \,, \quad 
p_i = \big\{ \hat{N}^{AB} , \,\hat T^{AB} , \,T , \,\PP_A \big\} ~.
\end{align}
We now wish to assess the physical content of each pair. 
In the $q_i$ we have two symmetric trace-free tensors: the shear and the part of the celestial metric that contributes to the trace-free part of $\delta\gamma^{0}_{AB}$ (the volume associated with this metric does not contribute to $\hat T^{AB} \delta\gamma_{AB}^0$), a scalar, and a vector. Their conjugate variables are respectively the news $\hat{N}^{AB}$, a trace free tensor $ \hat T^{AB}$, the scalar $T$, and the angular momentum $\PP_{A}$. From the expression \eqref{eq:Ttrace} we see that information about the mass is carried by the scalar $T$. The first novelty is that $\hat T_{AB}$ includes information about the source of log terms $\LL_{AB}$ due to the time dependence of the celestial metric. This was also noted in \cite{Riello:2024uvs}. If the volume associated with this metric is independent of $u$, $\partial_{u}\sqrt{\gamma_0} = 0$, then $\LL_{AB}$ completely drops out the symplectic potential (however see \cite{Geiller:2024ryw} for a recent discussion of the role played by $\LL_{AB}$ in gravity). Notice that this term cannot be recovered in the $\Lambda \to 0$ limit of the $\Lambda \neq 0$ result, since $\LL_{AB} = 0$ when $\Lambda \neq 0$.  It is interesting that the boundary source $\beta_0$ is not independent but appears together with the corner volume element. One might investigate whether a further relaxation of the gauge leads to a distinct symplectic pair involving $\beta_0$, as in three dimensions \cite{Geiller:2021vpg,Ciambelli:2023ott}.

The second new result is the presence of corner symplectic pairs on $\partial\BB$. They involve a set of $q_i$ that also appear in the symplectic pairs on $\BB$, but are conjugate to different quantities:
\begin{align}
&q^i_c=\{\gamma_{AB}^0, \ln\sqrt{\gamma_0}-4\beta_0 \}  \,, \quad p_i^c=\{\hat{\tau}^{AB}, \tau \} ~.
\end{align}
The $p_{i}^c$ appear in partial Bondi gauge, where there are no conditions on the traces $\gamma_1$ and $\gamma_2$. For instance in BS gauge $\gamma_1=0$, $\gamma_2=\frac12 \hat{\gamma}_{1}^{AB}\,\hat{\gamma}^{1}_{AB}$ so that both corner terms vanish. In PBG and for a boundary with corners, as is the case for future null infinity, these terms give a non-trivial contribution to the symplectic potential. At the level of the charges \cite{Geiller:2024amx}, it was shown that $\gamma_1,\gamma_2$ were associated to new symmetries parameters and had non-zero charges. However, unlike the mass or the angular momentum, the time dependence of these new symmetries and charges is not determined by the equations of motion. This is similar to what happens in three dimensions where these unconstrained quantities appear only at the corner $\partial \BB$ \cite{Geiller:2021vpg,Ciambelli:2023ott}.

Relaxing the condition $\UU_{0}^{A} = 0$ introduces a new symplectic pair $\PP_{A}\,\delta\UU_{0}^{A}$ on $\BB$, but does not appear to generate a new pair on $\partial \BB$. Surprisingly, including non-zero $\beta_0$ does not introduce any new pairs. Instead, $\beta_0 \neq 0$ merely dresses or shifts the expressions for some of the $p_i$ and $q^{i}$.

\subsubsection*{Symplectic pairs for $\Lambda \neq 0$}

When $\Lambda \neq 0$, the equation of motion \eqref{eq:LeadingEAB} gives the shear $\hat{\gamma}^{1}_{AB}$ in terms of $\gamma^{0}_{AB}$, $\UU_{0}^{A}$, and $\beta_0$. This reduces the number of symplectic pairs, since $\hat{\gamma}^{1}_{AB}$ is no longer independent. Re-writing the first term in \eqref{eq:finitesymplpot} using \eqref{eq:LeadingEAB}, we now have
\begin{align}
& q^i=\big\{\gamma_{AB}^0, \, \ln\sqrt{\gamma_0}-4\beta_0 , \,\UU_0^A \big\} \,, \quad 
p_i = \big\{\hat T_{(\Lambda)}^{AB} , \,T_{(\Lambda)} , \,\PP_{(\Lambda)}^{A} \big\} 
\end{align}
on $\BB$, and 
\begin{align}
&q^i_c=\{\gamma_{AB}^0, \ln\sqrt{\gamma_0}-4\beta_0 \}  \,, \quad p_i^c=\{\hat{\tau}_{(\Lambda)}^{AB}, \tau_{(\Lambda)} \} 
\end{align}
on $\partial \BB$. Notice that the process of rewriting the first term in \eqref{eq:finitesymplpot} preserves the combination $\ln\sqrt{\gamma_0}-4\,\beta_0$ appearing in the pairs. It also generates new contributions to the $p_{i}$ -- the subscript ${(\Lambda)}$ indicates quantities that differ from the expressions given above. Making contact between these symplectic pairs and familiar results for AlAdS and AldS spacetimes is a delicate procedure which will be discussed in the upcoming publication \cite{McNees:2025acf}.

\section{Discussion}
\label{sec:Discussion}
The central problem addressed in this paper is that direct variation of a bulk Lagrangian may lead to a symplectic potential $\Theta$ and current $\omega$ that are not compatible with the usual construction of charges at infinity. The limit \eqref{eq:ChargesWithLimit} does not exist, because the relevant component of the symplectic current (contracted with an asymptotic symmetry) does not fall off fast enough in \eqref{eq:Obstruction}.  This is the case for the usual form of the potential obtained from the Einstein-Hilbert Lagrangian \eqref{eq:EHBulk}. Like the analogous problem in the Hamiltonian formulation \cite{Regge:1974zd}, this issue is present even when the fields satisfy physically reasonable fall off conditions.

Expanding on earlier work \cite{McNees:2023tus}, we use the ambiguity \eqref{eq:BasicEquivalence} in the definition of $\Theta$ to address this problem, shifting to a new $\widetilde{\Theta} = \Theta + \dd\vartheta$ and $\tilde{\omega}$ such that the $r \to \infty$ limit exists in \eqref{eq:ChargesWithLimit}. The part of the shift \eqref{eq:varthetapresc} that removes this obstruction is uniquely determined by the bulk Lagrangian. In the example we focus on, this is the part of $\vartheta^{ur}$ that diverges as $r \to \infty$. There is, however, a residual ambiguity in the finite part of $\vartheta^{ur}$, which is \textit{not} determined by this condition. We make a specific choice for this part of the shift for Einstein-Hilbert gravity, with or without a cosmological constant. In partial Bondi gauge with a cosmological constant \cite{McNees:2025acf}, this choice yields the same symplectic current one would obtain when applying Comp\`ere-Marolf prescription \cite{Compere:2008us}\footnote{Comp\`ere-Marolf assume that the asymptotic boundary has no future or past component, therefore our results agree up to corner terms.}.
Our prescription for the symplectic potential at infinity gives a result that is finite for gravity with $\Lambda = 0$, and finite up to divergent $\delta$-exact terms when $\Lambda \neq 0$. In the latter case, the divergences have no effect on the symplectic current or charges, and can be removed from the first variation of the action with appropriate surface and corner terms.

An expression for the finite part of $\widetilde{\Theta}$ in PBG is given in \eqref{eq:finitesymplpot} in terms of symplectic pairs $p_{i}\,\delta q^{i}$ and $\delta$-exact parts. Our setup allows for a component $\BB$ of infinity with corner(s) $\partial \BB$, and the corner supports its own symplectic pairs. There are no restrictions on the time-dependence of the sources on $\BB$ and $\partial \BB$, and general variations of the transverse metric are allowed. An important point is that our choice for the finite part of the shift includes a part that is not $\delta$-exact, and hence contributes to the symplectic pairs that appear on $\partial \BB$.

The approach we follow here should not be confused with holographic renormalization via additional surface and corner terms in the action. Unless one imposes additional conditions on the fields at infinity (for instance, by viewing the corner ambiguity $\vartheta$ as corner symplectic potentials associated to these Lagrangians \cite{Compere:2008us}), these terms in the action change $\Theta$ by a $\delta$-exact term that does not affect $\omega$. 
Instead, our construction is compatible with any choice of boundary conditions on the fields. One should decide on boundary conditions for the theory, then add surface and corner terms to the action so that $\widetilde{\Theta}$ plus the variation of these new terms is consistent with that choice. An upcoming paper \cite{McNees:2025acf} will provide examples and recover familiar results for AlAdS spacetimes.

Another important feature of our method is that the fields need not be fully on-shell. Constraints associated with gauge fixing certain components of the metric \eqref{eq:PartialBondiMetric} have not been enforced. Instead, we impose a subset of the equations sufficient to determine the behavior of the fields at large $r$. A modification of the usual covariant phase space approach that accommodates this weaker set of conditions on the fields is described in appendix \ref{app:DiffeoCharges}. One can, of course, enforce the remaining equations, which describe flux balance laws for the mass and momentum. But relaxing the constraints leads to new contributions to the codimension-2 form used to compute charges, and in lower dimensional examples these contributions are known to support additional charges that are only visible off-shell \cite{Grumiller:2017qao}.

In this work, we established the content of symplectic pairs in PBG with or without a cosmological constant. The expression for the potential in section \ref{sec:finitesympl} includes the usual pairs on $\BB$ identified in Bondi-Sachs coordinates (generalized to the full set of boundary data consistent with PBG) as well as two new pairs with support on the corner $\partial \BB$. The pairs include contributions which are not visible when there are additional restrictions on the sources. For example, when $\Lambda = 0$ the traceless tensor $\hat{T}^{AB}$ that couples to $\delta\gamma^{0}_{AB}$ has a term proportional to $\LL_{AB}$ which is only present when the transverse metric $\gamma^{0}_{AB}$ has non-vanishing expansion: $\partial_{u}\sqrt{\gamma_0} \neq 0$. The new corner pairs involve two traces of terms in the large-$r$ expansion of $\gamma_{AB}$: $\gamma_1$ and $\gamma_2$, as well as the shear $\hat{\gamma}^{1}_{AB}$. From the analysis of the symmetry of the solution space \cite{Geiller:2022vto}, we know that there are symmetry generators associated to $\gamma_1$ and $\gamma_2$. Therefore, it would be interesting to investigate how each element of the symplectic potential transforms under the asymptotic symmetries. We could then examine the algebra of fluxes and charges, like the analysis in \cite{Geiller:2024amx} (which focused on charges in PBG with $\Lambda=0$ and some additional constraints on the boundary data). It would also be interesting to study this gauge using the BRST approach, as in \cite{Baulieu:2024oql}.

For gravity with vanishing cosmological constant, the PBG probes future null infinity which has two corners. It would be interesting to transform from PBG to coordinates adapted to these corners, in order to study the role of the new symplectic pairs. At spatial infinity they could be related to similar quantities introduced in \cite{Fiorucci:2024ndw} to describe superrotations. When approaching the corners of future null infinity, the fall-offs in $u$ become important. Typically, the fall-off of the shear is prescribed according to some physical process in the bulk, and then the fall-offs of other quantities (such as the mass) are determined by flux balance laws. But the quantities appearing in the corner symplectic pairs do not have associated flux balance laws, so understanding their fall off presents an interesting puzzle.

\section*{Acknowledgments}
CZ thanks Marc Geiller for collaboration on related topics. CZ thanks the participants of workshop Carrollian Physics and Holography hosted the Erwin Schr\"{o}dinger Institute in Vienna for discussions on this upcoming work, in particular Adrien Fiorucci and Romain Ruzziconi. We also thank Daniel Grumiller and Florian Ecker for stimulating conversations, and TU Wien for hospitality and support during the initial stages of this project. Research at Perimeter Institute is supported in part by the Government of Canada through the Department of Innovation, Science and Economic Development Canada and by the Province of Ontario through the Ministry of Colleges and Universities.

\appendix

\section{Codimension-2 form from diffeomorphism invariance}
\label{app:DiffeoCharges}

In this appendix we generalize covariant phase space methods \cite{GAWEDZKI1972307, Kijowski:1973gi, Kijowski:1976ze, Crnkovic:1986ex, Ashtekar:1990gc, Lee:1990nz, Wald:1993nt, Iyer:1994ys, Wald:1999wa, Barnich:2001jy, Barnich:2003xg, Barnich:2007bf} to field configurations which satisfy the partially on-shell conditions. The current $\omega$ is not conserved in that case, but diffeomorphism invariance implies the existence of a related current which is conserved and leads to a family of codimension-2 forms. Additional details may be found in section 2.2 of \cite{McNees:2023tus}. The construction here, which focuses on asymptotic symmetries generated by diffeomorphisms, can be extended to other gauge symmetries \cite{Compere:2007az} and anomalous transformations \cite{Chandrasekaran:2020wwn, Chandrasekaran:2021vyu, Freidel:2021cjp, Adami:2021sko, Odak:2023pga}.  

Consider a diffeomorphism-invariant theory with a set of tensor fields $\Psi_i$ described by a bulk Lagrangian $L_{M}$. The response of $L_{M}$ to a variation of the fields is
\begin{gather}\label{eq:AppendixBulkVariation}
    \delta L_{M} = E^{i}(\Psi)\,\delta\Psi_{i} + \partial_{\mu}\Theta^{\mu}(\Psi; \delta\Psi) ~.
\end{gather}
The tensor densities $E^{i}$ are the equations of motion, and $\Theta^{\mu}$ is the presymplectic potential. Anti-symmetrizing a second variation leads to a (pre)symplectic current
\begin{gather}
    \omega^{\mu}(\Psi;\delta_{1}\Psi, \delta_{2}\Psi) = \delta_{2} \Theta^{\mu}(\Psi; \delta_{1} \Psi) - \delta_{1} \Theta^{\mu}(\Psi; \delta_{2} \Psi) ~.
\end{gather}
Without loss of generality we can assume that these variations commute, $[\delta_2,\delta_1] L_{M} = 0 $, which implies
\begin{gather}\label{eq:CurrentDivergence1}
    \partial_{\mu}\omega^{\mu} + \delta_{2} E^{i}\,\delta_{1}\Psi_i - \delta_{1} E^{i}\,\delta_{2}\Psi_i = 0 ~. 
\end{gather}
This condition holds for all field configurations, regardless of whether or not the fields or their variations satisfy the equations of motion. When the field variations are on-shell, the current $\omega$ is conserved and can be written as $\omega = \dd k$. The codimension-2 form $k$, contracted with an asymptotic symmetry, is then used in the definition of the charges.

The construction above can be generalized to field configurations which are not fully on-shell, using diffeomorphism invariance. Under a diffeomorphism $\xi$ the fields $\Psi_i$ change by their Lie derivatives: $\delta_{\xi}\Psi_i = \pounds_{\xi}\Psi_i$. In that case, \eqref{eq:AppendixBulkVariation} can be integrated by parts to give
\begin{gather}\label{eq:DiffeoBulkLagrangian}
    \delta_{\xi} L_{M} = \xi^{\mu} N_{\mu} + \partial_{\mu}\Big(\Theta^{\mu} + S_{\xi}^{\mu}\Big) ~.
\end{gather}
The quantity $N_\mu$ is a combination of the fields, their derivatives, and the equations of motion which vanishes identically as a consequence of diffeomorphism invariance
\begin{gather}
    N_{\mu} = E^{i}\partial_{\mu}\Psi_i - \partial_{\nu}\big(E^i \Psi_i\big)^{\mu}{}_{\nu}  = 0 ~.
\end{gather}
Here, $\big(E^i \Psi_i\big)^{\mu}{}_{\nu}$ are contractions of the tensor densities $E^{i}$ and tensors $\Psi_i$ appearing in
\begin{gather}
   E^{i} \delta_{\xi}\Psi_i = E^{i}\,\xi^{\mu} \partial_{\mu}\Psi_i + \big(E^i \Psi_i\big)^{\mu}{}_{\nu} \partial_{\mu}\xi^{\nu} ~.
\end{gather}
The current $S_{\xi}^{\mu}$ appearing as a divergence in \eqref{eq:DiffeoBulkLagrangian} is 
\begin{gather}
    S_{\xi}^{\mu} = \xi^{\nu}\,\big(E^i \Psi_i\big)^{\mu}{}_{\nu} ~.
\end{gather}
This is sometimes referred to as the ``weakly vanishing Noether current,'' since it is zero when the fields are on-shell. But unlike $N_{\mu}$, which vanishes for any field configuration, $S_{\xi}^{\mu}$ may have non-zero components if the fields do not satisfy all the equations of motion. This will be the case for the partially on-shell conditions used in this paper.

Since $N_{\mu}=0$ by diffeomorphism invariance, independent of the equations of motion for the fields and their variations, its linearization around any field configuration  must also vanish: $\delta N_{\mu} = N_{\mu}(\Psi + \delta \Psi) - N_{\mu} (\Psi) = 0$. This implies
\begin{gather}
    \delta N_{\mu} = \delta E^{i}\,\partial_{\mu} \Psi_{i} + E^{i} \partial_{\mu}\delta\Psi_i - \partial_{\nu} \delta\big(E^{i}\Psi_i\big)^{\mu}{}_{\nu} = 0 ~.
\end{gather}
Contracting this with the diffeomorphism $\xi^{\mu}$, and using the fact that the $E^i$ transform as tensor densities, we obtain
\begin{gather}
    \delta E^{i} \delta_{\xi}\Psi_i - \delta_{\xi}E^{i} \delta\Psi_i = \partial_{\mu}\Big( \xi^{\nu}\,\delta \big(E^{i}\Psi_i\big)^{\mu}{}_{\nu} - \xi^{\mu} E^{i}\,\delta\Psi_i \Big) ~.
\end{gather}
The left-hand side is precisely the combination of terms appearing in \eqref{eq:CurrentDivergence1}, when $\delta_2$ is an arbitrary field variation and $\delta_1$ is a field variation that takes the same form $\delta_{\xi}$ as a diffeomorphism. Thus, independent of the equations of motion, all field configurations satisfy
\begin{gather}
    \partial_{\mu}\Big(\omega_{\xi}^{\mu} + \xi^{\nu}\,\delta \big(E^{i}\Psi_i\big)^{\mu}{}_{\nu} - \xi^{\mu} E^{i}\,\delta\Psi_i \Big)  = 0 ~,
\end{gather}
with $\omega_{\xi}^{\mu} \defeq \omega^{\mu}(\Psi;\delta_{\xi}\Psi, \delta \Psi)$. This is a consequence of diffeomorphism invariance. Therefore, it is always possible to find codimension-2 forms $k_{\xi}^{\mu\nu}$ such that\,\footnote{A similar result is arrived at via a different approach in \cite{Fiorucci:2021pha}.}
\begin{gather}\label{eq:Codim2Defintion1}
    \partial_{\nu}k_{\xi}^{\mu\nu} = \omega_{\xi}^{\mu} + \xi^{\nu}\,\delta \big(E^{i}\Psi_i\big)^{\mu}{}_{\nu} - \xi^{\mu} E^{i}\,\delta\Psi_i ~.
\end{gather}
When the fields and field variations are fully on-shell, $E^i = \delta E^{i} = 0$, this reduces to the usual definition \cite{Iyer:1994ys, Wald:1999wa}.

Rather than enforcing all equations of motion, we impose the milder condition $E^{i}\delta\Psi_i = 0$.\,\footnote{To be precise, let $a$ be an index labeling the fields that are not fixed by the choice of gauge. Then we enforce $E^{a} = \delta E^{a} = 0$.} As a result, the last term in \eqref{eq:Codim2Defintion1} vanishes, leaving
\begin{gather}\label{eq:Codim2Definition2}
    \partial_{\nu}k_{\xi}^{\mu\nu} = \omega_{\xi}^{\mu} + \xi^{\nu}\,\delta \big(E^{i}\Psi_i\big)^{\mu}{}_{\nu} ~.     
\end{gather}
When the weakly vanishing Noether current does not vanish, mixing between non-zero components of $E^i$ and $\Psi_i$ (and their variations) may support charges which are not visible when $E^i = \delta E^i = 0$. This occurs, for example, in certain lower-dimensional theories of gravity where a holographic dual exhibits extended off-shell symmetries that are broken on-shell \cite{Grumiller:2017qao}.

The arguments given in section \ref{sec:GeneralConstruction} still apply to the generalization described above. With this definition of $k_{\xi}$, it is the $u$ component of the right-hand side of \eqref{eq:Codim2Definition2} which must fall off as $o(r^{-1})$ so that the $r \to \infty$ limit exists in the definition of $\delta\!\!\!/Q_{\xi}$. For Einstein gravity with the partially on-shell conditions given in section \ref{sec:PartialBondiGauge},
the $u$ component of the extra term vanishes since
\begin{gather}
    \big(E^{i}\Psi_i\big)^{u}{}_{\nu} = 2\,E^{u\lambda}\,g_{\lambda\nu} = 0~.
\end{gather}
Therefore our prescription for the potential, which shifts $\omega^{u} \to \tilde{\omega}^{u} = o(r^{-1})$, is sufficient for charges constructed using the generalization \eqref{eq:Codim2Definition2} of the codimension-2 form. However, the $r$ component of the extra term will in general be non-zero:
\begin{gather}
    \xi^{\nu} \delta \big(E^{i}\Psi_i\big)^{r}{}_{\nu} = 2\,\xi^{u}\,\delta \big( E^{rr}\,g_{ru} + E^{rA}\,g_{Au}\big) + 2\,\xi^{B}\,\delta \big(E^{rA}\,g_{AB}\big) ~.
\end{gather}
As a result, charges built from $\tilde{k}_{\xi}^{ru}$ receive a contribution that is not present when the fields are fully on-shell.

\section{Solution space in partial Bondi gauge}\label{app:EOMPBG}

In this appendix we give detailed results for all quantities in \eqref{eq:gammaAB}-\eqref{expUbetaV}. When solving a particular component of $E^{\mu\nu}$ in the following subsections, it is assumed that the previous components of $E^{\mu\nu}$ have already been set to zero. So, for instance, a term proportional to $R_{rr}$ in $E^{uA}$ does not appear since we already assume that $E^{uu} = 0$ holds. In addition, some expressions have been simplified using identities for the contractions of symmetric, traceless tensors in two dimensions. These identities are reviewed in appendix \ref{app:Identities}.

\subsection*{The $uu$ component}
In terms of the four dimensional Ricci tensor, the $uu$ component of $E^{\mu\nu}$ is
\begin{gather}
    2\kappa^{2}\,E^{uu} = - e^{-2\beta}\sqrt{\gamma}\,R_{rr} = 0 ~.
\end{gather}  
This equation does not depend on $\Lambda$. It fixes the coefficients in the large $r$ expansion of $\beta$, with an integration constant appearing at leading order.
	\begin{align}
		\beta_0 = &\,\, \text{Free} \\
        \beta_{1} =&\,\, 0 \\
  		\beta_{2} = & \,\, \frac{1}{32}\hat{\gamma}_{1}^{AB}\,\hat{\gamma}^{1}_{AB} + \frac{1}{64}\,\big(\gamma_{1}\big)^{2} - \frac{1}{8}\,\gamma_{2} \\
        \beta_{3} = & \,\, \frac{1}{8}\,\AA_{3} + \frac{1}{6}\,\LL^{AB}\,\hat{\gamma}^{1}_{AB} - \frac{1}{64}\,\gamma_{1} \,\hat{\gamma}_{1}^{AB} \, \hat{\gamma}^{1}_{AB} - \frac{1}{128}\,\big(\gamma_{1}\big)^{3} + \frac{1}{16}\,\gamma_1\,\gamma_2 - \frac{1}{4}\,\gamma_{3} \\
        \tilde{\beta}_{3} = &\,\,-\frac{1}{4}\,\AA_{3} ~.
	\end{align}
 The quantity $\LL_{AB}$ appearing in the expression for $\beta_3$ was introduced in \eqref{eq:LABDefinition}. It is given by
 \begin{gather}
    \LL_{AB} = \hat{\gamma}^{2}_{AB} - \frac{1}{4}\,\gamma_{1}\,\hat{\gamma}^{1}_{AB} ~.
\end{gather}

\subsection*{The $uA$ component}
After imposing $E^{uu}=0$, the $uA$ component of $E^{\mu\nu}$ reduces to
\begin{gather}
    2\kappa^{2}\,E^{uA} = \sqrt{\gamma}\,\gamma^{AB}\,R_{rB} = 0 ~.
\end{gather}  
This equation is independent of $\Lambda$. It fixes the coefficients in the large $r$ expansion of $\UU^{A}$, with integration constants appearing at leading order ($r^0$) and $\OO(r^{-3})$.
	\begin{align}
		\UU_{0}^{A} = &\,\, \text{Free} \\
		\UU_{1}^{A} = &\,\, \DD^{A}(e^{2\beta_0}) \\
  		\UU_{2}^{A} = &\,\, -\frac{1}{2}\,\DD_{B}\big(e^{2\beta_0} \hat{\gamma}_{1}^{AB}\big) + \frac{1}{4}\,e^{2\beta_{0}} \DD^{A}\gamma_{1} - \frac{1}{4}\,\gamma_{1}\,\DD^{A}\big(e^{2\beta_0}\big)\\
		\UU_{3}^{A} = &\,\,\text{Free} \\
		\tilde{\UU}_{3}^{A} = & \,\, -\frac{2}{3}\,e^{2\beta_{0}}\,\DD_{B}\LL^{AB} ~.
	\end{align}

\subsection*{The $ur$ component}
After imposing the previous equations, the remaining part of the $ur$ component of $E^{\mu\nu}$ depends on $\Lambda$
\begin{gather}
2\kappa^{2}\,E^{ur} = - \frac{1}{2}\,\sqrt{\gamma}\,\Big(\gamma^{AB}\,R_{AB} - 2\Lambda\Big) = 0 ~.    
\end{gather}
Note that $R_{AB}$ are components of the full, four-dimensional Ricci tensor. Solving this equation fixes coefficients in the large $r$ expansion of $\VV$, with a single integration constant appearing at order $\OO(r^{-1})$.\,\footnote{Recall that the leading term in \eqref{eq:VExpansion} is $\OO(r^2)$, so the integration constant appears at the third sub-leading order relative to the leading term.} 
\begin{align} \label{eq:V0eqn}
			\VV_{0} = & \,\, \frac{\Lambda}{3}\,e^{2\beta_0} \\ \label{eq:V1eqn}
   			\VV_{1} = & \,\, \frac{1}{2}\,\VV_{0}\,\gamma_{1} - \partial_{u} \ln \sqrt{\gamma_{0}} - \DD_{A} \UU_{0}^{A} \\ \label{eq:V2eqn}
   			\VV_{2} = &\,\, - \frac{1}{4}\,\gamma_{1}\,\VV_{1} - \frac{1}{2\,\sqrt{\gamma_0}}\,\partial_{u}\Big(\sqrt{\gamma_0}\,\gamma_{1}\Big) - \frac{1}{2}\,\DD_{A}\Big(\gamma_{1}\,\UU_{0}^{A}\Big)  - \frac{1}{2}\,e^{2\beta_0}\,\RR - \DD^{2}\Big( e^{2\beta_0} \Big)\\ \nonumber
            & \,\, + \frac{3}{4}\,\VV_0\,\left(\gamma_{2} + \frac{1}{8}\,\big(\gamma_{1}\big)^{2} - \frac{3}{4}\, \hat{\gamma}_{1}^{AB} \hat{\gamma}^{1}_{AB}\right) \\ \label{eq:V3eqn}
			\VV_{3} = & \,\, \text{free} \\ \label{eq:V3logeqn}
			\tilde{\VV}_{3} = & \,\, \VV_{0}\,\left( \AA_{3} - \frac{1}{2}\,\LL^{AB}\,\hat{\gamma}^{1}_{AB} \right) ~.
\end{align}
There are no subtleties for vanishing cosmological constant. The solution in that case is obtained by setting $\Lambda = 0$ in the expressions above. Then $\VV_0$ vanishes, which gives a degenerate result for the unphysical line element $ds^{2} / r^{2}$ in the $r \to \infty$ limit. This also causes the log term $\tilde{V}_{3}$ to vanish, along with some terms proportional to $\gamma_1$ in $\VV_1$ and $\VV_2$.

\subsection*{The $AB$ component}
The $AB$ component of $E^{\mu\nu}$ can be split into trace and trace-free parts with respect to $g_{AB} = \gamma_{AB}$. The trace part depends on $\Lambda$
\begin{align}\label{eq:TraceEAB}
    2\,\kappa^{2}\,E^{AB}\,\gamma_{AB} = &\,\, -2\,\sqrt{\gamma}\,\Big(R_{ru} + e^{2\beta}\,\Lambda \Big) ~,
\end{align}
while the trace-free part is independent of $\Lambda$
\begin{align}\label{eq:TraceFreeEAB}
    2\,\kappa^{2}\,E^{CD}\,\Big(\delta^{A}{}_{C}\,\delta^{B}{}_{D} - \frac{1}{2}\,\gamma^{AB}\,\gamma_{CD} \Big) = & \,\, - e^{2\beta} \, \sqrt{\gamma} \, \Big( \gamma^{AC} \, \gamma^{BD} - \frac{1}{2}\,\gamma^{AB}\,\gamma^{CD} \Big)\,R_{CD} ~.
\end{align}
As before, $R_{CD}$ refers to components of the four dimensional Ricci tensor -- the trace-free equation is not related to the contracted Bianchi identity in two dimensions. 

Both the trace and trace-free parts of $E^{AB}$ give equivalent results once the $\Lambda$-dependent equation $E^{ur} = 0$ is imposed. Here we focus on the trace-free part \eqref{eq:TraceFreeEAB}, since the analysis is simpler. First we present the conditions obtained at each order without any reference to the other equations of motion. In particular, we make no assumptions about $\Lambda$. The only simplifications in the sub-leading orders comes from the equations that appear at preceding (higher) orders in $E^{AB}$.
\begin{itemize}[topsep=0.5em, itemsep=1.5em]
    \item $\OO(r^3)$
    \begin{gather}\label{eq:Or3Traceless}
        \partial_u \gamma^{0}_{AB} - \gamma^{0}_{AB}\,\partial_{u}\ln\sqrt{\gamma_{0}} + 2 \, \DD_{\langle A} \UU^{0}_{B\rangle} = \VV_0\,\hat{\gamma}^{1}_{AB} 
    \end{gather}
    \item $\OO(r^2)$
    \begin{gather}\label{eq:Or2Traceless}
        0 = \VV_{0}\,\LL_{AB} + \frac{1}{2}\,\hat{\gamma}^{1}_{AB} \,\left( \partial_{u} \ln\sqrt{\gamma_{0}} + \DD_{C}\UU_{0}^{C} - \frac{1}{2}\,\VV_{0}\,\gamma_{1} + \VV_{1} \right)
    \end{gather}    
    \item $\OO(r^1)$
    \begin{align}\label{eq:Or1Traceless}
        0 = &\,\, \left(\delta_{A}{}^{C} \delta_{B}{}^{D} - \frac{1}{2}\,\gamma^{0}_{AB}\,\gamma_{0}^{CD} \right)\,\left[\partial_{u} \LL_{CD} + \pounds_{\UU_{0}} \LL_{CD} + \frac{3}{2}\,\VV_{0}\,\AA^{3}_{CD} \right]
    \end{align}
\end{itemize}
In this last equation, the second term in square brackets is the Lie derivative of $\LL_{CD}$ along $\UU_{0}^{A}$. 

The appearance of $\VV_0$ and $\VV_1$ in the expressions above leads to qualitatively different results for vanishing and non-zero cosmological constant. When $\Lambda \neq 0$, applying the results obtained from the $E^{ur}$ equation gives
	\begin{gather} \label{eq:EABwithCC1}
		\partial_{u} \gamma^{0}_{AB} - \gamma^{0}_{AB} \partial_{u} \ln \sqrt{\gamma_{0}} + 2\,\DD_{\langle A} \UU^{0}_{B\rangle} = \frac{\Lambda}{3}\,e^{2\beta_{0}}\,\hat{\gamma}^{1}_{AB} \\ \label{eq:LABwithCC}
		\LL^{AB} = 0 \\ \label{eq:hatA3ABwithCC}
        \hat{\AA}^{3}_{AB} = 0 ~.
	\end{gather}
In this case the shear is constrained by the flow of the two dimensional metric $\gamma^{0}_{AB}$ in the $u$ and $\UU_{0}^{A}$ directions, and both $\LL_{AB}$ and the traceless part of $\AA^{3}_{AB}$ vanish. However, when $\Lambda = 0$, $\VV_0$ vanishes and we instead find the conditions
	\begin{gather}
		\partial_{u} \gamma^{0}_{AB} - \gamma^{0}_{AB} \partial_{u} \ln \sqrt{\gamma_{0}} + 2\,\DD_{\langle A} \UU^{0}_{B\rangle}  = 0 \\ \label{eq:EABnoCC1}
        \partial_{u} \LL_{AB} +  \pounds_{\UU_0}\LL_{AB} = 0 ~.
	\end{gather}
 In this case there is no condition on the shear $\hat{\gamma}^{1}_{AB}$. Likewise, $\LL_{AB}$ is no longer forced to vanish. Instead, it satisfies a differential condition relating its Lie derivatives in the $u$ and $\UU_{0}^{A}$ directions. There is a similar condition on the traceless part of $\partial_{u}\AA^{3}_{AB}$ that comes from the $\OO(\ln r)$ term in $E^{AB}$, but it is not needed for our analysis.

\section{Useful identities}\label{app:Identities}
In this appendix we collect various identities and results used throughout the text. These fall into two categories: contractions of two-dimensional tensors and their derivatives that vanish identically, and quantities that can be rewritten in useful ways via integration-by-parts and applications of the equations of motion.

The first group of identities are specific to two dimensions, where the generalized Kronecker delta with six indices vanishes 
\begin{gather}\label{eq:GenKronecker}
    \delta_{ABC}^{DEF} = 3!\,\delta_{[A}{}^{D}\,\delta_{B}{}^{E}\,\delta_{C]}{}^{F} = 0 ~.
\end{gather}
Contracting this with various tensors leads to quantities which are zero, even though this may not be obvious from the expressions themselves. Examples of such identities include
\begin{gather} \label{eq:TwoTensorIdentity}
    G_{A}{}^{C}\,H_{BC} + G_{B}{}^{C}\,H_{AC} - \gamma_{AB}\,G^{CD}\,H_{CD} = 0 \vphantom{\frac{1}{2}}
    \\ \label{eq:TwoTebsorDiffIdentity} 
    G_{A}{}^{C}\,\partial H_{C}{}^{B} + G_{C}{}^{B}\,\partial H_{A}{}^{C} - \delta_{A}{}^{B}\,G_{C}{}^{D}\,\partial H_{D}{}^{C} = 0 \vphantom{\frac{1}{2}}
    \\ \label{eq:SimpleTwoTensorCovDIdentity}
    G_{AC}\,D_{B} H^{CB} + G^{CB}\,D_{B} H_{AC} - G^{CD}\,D_{A} H_{CD} = 0 \vphantom{\frac{1}{2}}
    \\ \label{eq:ThreeTensorIdentityt}
    G_{A}{}^{C}\,H_{C}{}^{D}\,F_{DB} - \frac{1}{2}\,G_{AB}\,H^{CD} F_{CD} + \frac{1}{2}\,H_{AB}\,G^{CD} F_{CD} - \frac{1}{2}\,F_{AB}\,G^{CD} H_{CD} = 0 ~.
\end{gather}
where $G_{AB}$, $H_{AB}$, $F_{AB}$ are symmetric, traceless tensors, $\partial$ is any differential operator such that $\partial H_{A}{}^{B}$ remains symmetric and traceless, and $D_{A}$ is the covariant derivative compatible with the two-dimensional metric $\gamma_{AB}$. The first identity with $G_{AB} = H_{AB} = \hat{\gamma}^{1}_{AB}$ is used extensively throughout this paper, to rewrite
\begin{gather}
    \hat{\gamma}^{1}_{A}{}^{C}\,\hat{\gamma}^{1}_{BC} = \frac{1}{2}\,\gamma^{0}_{AB}\,\hat{\gamma}_{1}^{CD}\,\hat{\gamma}^{1}_{CD} ~.
\end{gather}
Next, let $\partial$ be any differential operator that obeys the Leibniz rule. Then the identities above imply
\begin{multline}\label{eq:TwoTensorDiffIdentity2}
    \partial G_{AC}\,H_{B}{}^{C} + G_{A}{}^{C}\,\partial H_{BC} + \partial G_{BC}\,H_{A}{}^{C} + G_{B}{}^{C}\,\partial H_{AC} = \\
    = \gamma_{AB}\,G^{CD}\,\partial H_{CD} + \gamma_{AB}\,H^{CD}\,\partial G_{CD} + G_{AB}\,H^{CD} \partial\gamma_{CD} + H_{AB}\,G^{CD} \partial\gamma_{CD} ~.
\end{multline}    
This is especially useful when $\partial$ relates one of the quantities to an equation of motion. For example, when $G_{AB} = H_{AB} = \hat{\gamma}^{1}_{AB}$, and using the equation of motion for $\partial_{u}\gamma^{0}_{AB}$, we have
\begin{gather}
    \hat{\gamma}^{1}_{A}{}^{C} \partial_u \hat{\gamma}^{1}_{BC} + \hat{\gamma}^{1}_{A}{}^{C} \partial_u \hat{\gamma}^{1}_{BC} = \gamma^{0}_{AB}\,\hat{\gamma}_{1}^{CD} \partial_u \hat{\gamma}^{1}_{CD} + \VV_0\,\hat{\gamma}^{1}_{AB}\,\hat{\gamma}_{1}^{CD} \hat{\gamma}^{1}_{CD} ~.
\end{gather}
This is used in certain calculations to rewrite terms involving the news $\hat{N}_{AB}$. 

Another example of the sort of identity described above is the vanishing of the Einstein tensor in two dimensions
\begin{gather}
    \RR_{AB} - \frac{1}{2}\,\gamma_{AB}\,\RR = 0 ~.
\end{gather}
Since this is true for any $\gamma_{AB}$, one can linearize $\gamma_{AB} \to \gamma_{AB} + \delta\gamma_{AB}$ to obtain
\begin{multline}\label{eq:LinearBianchi}
    0 = D^{C}D_{A}\delta\gamma_{CB} + D^{C}D_{B}\delta\gamma_{AC}-\gamma_{AB} D^{C} D^{D} \delta\gamma_{CD} - D_{A}D_{B} (\delta\gamma)^{C}{}_{C}\\ - D^{2}\delta\gamma_{AB}
    + \gamma_{AB} D^{2}(\delta\gamma)^{C}{}_{C} -\RR\,\left(\delta\gamma_{AB} - \frac{1}{2}\,\gamma_{AB}\,(\delta\gamma)^{C}{}_{C} \right) ~.
\end{multline}
This is just a special case of the more general result obtained from \eqref{eq:TwoTensorIdentity} when $G_{AB}$ is the symmetric and traceless differential operator
\begin{gather}
    D_{A} D_{B} + D_{B} D_{A} - \gamma_{AB}\,D_{C} D^{C} ~,
\end{gather}
which gives
\begin{gather}
    D^{C} D_{A} H_{CB} +  D^{C} D_{B} H_{AC} - \gamma_{AB}\, D_{C} D_{D} H^{CD} - D^{C} D_{C}H_{AB} - \RR\,H_{AB} = 0 ~.
\end{gather}
Some terms that appear in the finite (as $r \to \infty$) part of $\Theta^{r}$ can be eliminating with this result, by replacing $H_{AB}$ with symmetric, traceless tensors constructed from the fields $\beta_n$, $\UU_{n}^{A}$, $\VV_n$, and $\gamma^{n}_{AB}$.

Other identities, built from variational and differential operators acting on the fields, are used in one or two instances to simplify an expression by reducing the number of distinct terms. For example, contracting \eqref{eq:GenKronecker} with the quantity $\DD_{D}\beta_{0}\,\DD_{C}\hat{\gamma}^{1}_{AB}$ leads to the result
\begin{gather}
      \DD^{C} \beta_0 \,\DD_{C}\hat{\gamma}^{1}_{AB} - \DD^{C}\beta_0 \,\DD_{\langle A} \hat{\gamma}^{1}_{B\rangle C} - \DD_{\langle A} \beta_0 \,\DD^{C}\hat{\gamma}^{1}_{B\rangle C}  = 0 ~,
\end{gather}
while contracting with $\hat{\gamma}^{1}_{AB} \, \DD_{C}\beta_{0} \, \DD_{D}\beta_{0}$ gives
\begin{gather}
    \gamma^{0}_{AB}\,\hat{\gamma}^{1}_{CD}\DD^{D}\beta_0 + \hat{\gamma}^{1}_{AB} \DD_{C}\beta_0 - \hat{\gamma}^{1}_{C \langle A}\DD_{B\rangle}\beta_{0} - \gamma^{0}_{C \langle A}\hat{\gamma}^{1}_{B\rangle D} \DD^{D}\beta_{0} = 0 ~.
\end{gather}
These and other identities are used to obtain expressions like \eqref{eq:ThatAB} for $\hat{T}^{AB}$.

Finally, the partially on-shell conditions can be used to replace certain terms in the large $r$ expansion of $\Theta^{r}$ with expressions that are equivalent up to total derivatives $\partial_{A}(\ldots)$. These total derivatives are discarded in integrals over $\BB$ and $\partial \BB$. An especially important example follows from linearizing \eqref{logeom} and applying the leading result \eqref{eq:LeadingEAB} of the $E^{AB}$ equation of motion. Then we obtain
\begin{gather}\label{eq:LogIdentity}
    \sqrt{\gamma}\,\delta\UU_{0}^{B}\,\DD^{A}\LL_{AB} = \partial_{u}\left(\frac{1}{2}\,\sqrt{\gamma_0}\,\LL^{AB}\,\delta\gamma^{0}_{AB}\right) + \partial_{A}\left(\frac{1}{2}\,\sqrt{\gamma_0}\,\UU_{0}^{A}\,\LL^{BC}\,\delta\gamma^{0}_{AB} + \sqrt{\gamma_0}\,\LL^{A}{}_{B}\,\delta\UU_{0}^{B} \right) ~.
\end{gather}
This result confirms that a $\ln r$ divergent corner term in $\Theta^{r}$ is canceled by a similar term in $\vartheta^{ur}$ after the shift $\Theta^{\mu} \to \Theta^{\mu} + \partial_{\nu}\vartheta^{\mu\nu}$. It is worth reiterating here that neglecting $\partial_{A}(\ldots)$ terms is purely for convenience. Keeping track of such terms in $\Theta$, one finds that the $\partial_{A}(\ldots)$ part of \eqref{eq:LogIdentity} participates in a similar cancellation with other terms which we have neglected.


\providecommand{\href}[2]{#2}\begingroup\raggedright\endgroup

\end{document}